\documentclass[preprints]{aastex61}

\usepackage{lipsum} 
\usepackage{footnote}
\usepackage{fancyhdr}
\shorttitle{Probing ISM Structure in Tr 14/Carina I Using STO2}
\shortauthors{Seo et al.}

\begin{document}

\title{Probing ISM Structure in Trumpler 14 \& Carina I Using The Stratospheric Terahertz Observatory 2}

\author{Young Min Seo}
\affiliation{Jet Propulsion Laboratory, California Institute of Technology, 4800 Oak Grove Drive, Pasadena, CA, 91109, USA}
\correspondingauthor{Young Min Seo}
\email{seo3919@gmail.com}

\author{Paul F. Goldsmith}
\affiliation{Jet Propulsion Laboratory, California Institute of Technology, 4800 Oak Grove Drive, Pasadena, CA, 91109, USA}

\author{Christopher K. Walker}
\affiliation{Department of Astronomy \& Steward Observatory, University of Arizona, 933 N. Cherry Ave., Tucson, AZ 85721, USA}

\author{David J. Hollenbach}
\affiliation{Carl Sagan Center, SETI Institute, 189 Bernado Avenue, Mountain View, CA 94043-5203, USA}

\author{Mark G. Wolfire}
\affiliation{Department of Astronomy, University of Maryland, College Park, MD 20742-2421, USA}

\author{Craig A. Kulesa}
\affiliation{Department of Astronomy \& Steward Observatory, University of Arizona, 933 N. Cherry Ave., Tucson, AZ 85721, USA}

\author{Volker Tolls}
\affiliation{Harvard–-Smithsonian Center for Astrophysics, 60 Garden Street, Cambridge, MA 02138, USA}

\author{Pietro N. Bernasconi}
\affiliation{Applied Physics Laboratory, Johns Hopkins University}

\author{\"Umit Kavak}
\affiliation{Kapteyn Astronomical Institute, University of Groningen, Netherlands}
\affiliation{SRON Netherlands Institute for Space Research, Landleven 12, 9747 AD Groningen, The Netherlands}

\author{Floris F.S. van der Tak}
\affiliation{Kapteyn Astronomical Institute, University of Groningen, Netherlands}
\affiliation{SRON Netherlands Institute for Space Research, Landleven 12, 9747 AD Groningen, The Netherlands}

\author{Russ Shipman}
\affiliation{SRON Netherlands Institute for Space Research, Landleven 12, 9747 AD Groningen, The Netherlands}

\author{Jian Rong Gao}
\affiliation{SRON Netherlands Institute for Space Research, Landleven 12, 9747 AD Groningen, The Netherlands}
\affiliation{Faculty of Applied Science, Delft University of Technology, Delft, The Netherlands}

\author{Alexander Tielens}
\affiliation{Leiden Observatory, Leiden University, P.O. Box 9513, NL-2300 RA Leiden, The Netherlands}

\author{Michael G. Burton}
\affiliation{School of Physics, University of New South Wales, Sydney, NSW, 2052, Australia}
\affiliation{Armagh Observatory and Planetarium, College Hill, Armagh, BT61 9DG, Northern Ireland, UK}

\author{Harold Yorke}
\affiliation{NASA Ames SOFIA Science Center, N211, Mountain View, CA 94043, USA}

\author{Erick Young}
\affiliation{NASA Ames SOFIA Science Center, N211, Mountain View, CA 94043, USA}

\author{William L. Peters}
\affiliation{Department of Astronomy \& Steward Observatory, University of Arizona, 933 N. Cherry Ave., Tucson, AZ 85721, USA}

\author{Abram Young}
\affiliation{Department of Astronomy \& Steward Observatory, University of Arizona, 933 N. Cherry Ave., Tucson, AZ 85721, USA}

\author{Christopher Groppi}
\affiliation{Department of Astronomy, Arizona State University, Tempe, AZ, USA}

\author{Kristina Davis}
\affiliation{Department of Astronomy, Arizona State University, Tempe, AZ, USA}
\affiliation{University of California Santa Barbara, Santa Barbara, CA, 93106, USA}

\author{Jorge L. Pineda}
\affiliation{Jet Propulsion Laboratory, California Institute of Technology, 4800 Oak Grove Drive, Pasadena, CA, 91109, USA}

\author{William D. Langer}
\affiliation{Jet Propulsion Laboratory, California Institute of Technology, 4800 Oak Grove Drive, Pasadena, CA, 91109, USA}

\author{Jonathan H. Kawamura}
\affiliation{Jet Propulsion Laboratory, California Institute of Technology, 4800 Oak Grove Drive, Pasadena, CA, 91109, USA}

\author{Antony Stark}
\affiliation{Harvard–-Smithsonian Center for Astrophysics, 60 Garden Street, Cambridge, MA 02138, USA}

\author{Gary Melnick}
\affiliation{Harvard–-Smithsonian Center for Astrophysics, 60 Garden Street, Cambridge, MA 02138, USA}

\author{David Rebolledo}
\affiliation{Joint ALMA Observatory, Alonso de Córdova 3107, Vitacura, Santiago, Chile}
\affiliation{National Radio Astronomy Observatory, 520 Edgemont Road, Charlottesville, VA 22903, USA }

\author{Graeme F. Wong}
\affiliation{School of Physics, University of New South Wales, Sydney, NSW, 2052, Australia}
\affiliation{School of Computing Engineering and Mathematics, Western Sydney University, Locked Bay 1797, Penrith, NSW 2751, Australia}

\author{Shinji Horiuchi}
\affiliation{CSIRO Astronomy and Space Science, Canberra Deep Space Communications Complex, PO Box 1035, Tuggeranong, ACT 2901, Australia}

\author{Thomas B. Kuiper}
\affiliation{Jet Propulsion Laboratory, California Institute of Technology, 4800 Oak Grove Drive, Pasadena, CA, 91109, USA}

\begin{abstract}

We present observations of the Trumpler 14/Carina I region carried out using the Stratospheric Terahertz Observatory 2 (STO2). The Trumpler 14/Carina I region is in the west part of the Carina Nebula Complex, which is one of the most extreme star-forming regions in the Milky Way. We observed Trumpler 14/Carina I in the 158 $\mu$m transition of [C\,{\sc ii}] with a spatial resolution of 48$''$ and a velocity resolution of 0.17 km s$^{-1}$. The observations cover a 0.25$^\circ$ by 0.28$^\circ$ area with central position {\it l} = 297.34$^\circ$, {\it b} = -0.60$^\circ$. The kinematics show that bright [C\,{\sc ii}] structures are spatially and spectrally correlated with the surfaces of CO clouds, tracing the photodissociation region and ionization front of each molecular cloud. { Along 7 lines of sight that traverse Tr 14 into the dark ridge to the southwest, we find that the [C\,{\sc ii}] luminosity from the HII region is 3.7 times that from the PDR. In same los we find in the PDRs an average ratio of 1:4.1:5.6 for the mass in atomic gas:dark-CO gas: molecular gas traced by CO. Comparing multiple gas tracers including HI 21cm, [C\,{\sc ii}], CO, and radio recombination lines, we find that the HII regions of the Carina Nebula Complex are well-described as HII regions with one-side freely expanding towards us, consistent with the champagne model of ionized gas evolution. The dispersal of the GMC in this region is dominated by EUV photoevaporation; the dispersal timescale is 20-30 Myr.}

\end{abstract}


\section{INTRODUCTION} \label{sec:intro}

The interstellar medium (ISM) is one of the main constituents of galaxies, and understanding its life cycle has been a fundamental issue for following galaxy evolution as well as star and planet formation. The ISM is observed to have multiple phases including hot/warm ionized gas, warm/cold neutral gas, and cold molecular gas \citep[see][for a more detailed classification]{snow06}. The ISM cycles through these phases through dynamic processes including cloud formation, star formation, stellar winds, and supernova explosions. In the ISM life cycle, the transition from diffuse atomic gas to dense molecular clouds and the destruction of molecular clouds to diffuse gas by stellar feedback may be { critical steps associated with star formation} that may control the rate of star formation in galaxies. However, the { ISM life cycle} is still poorly understood because the transitions between the ISM phases go through multiple complex processes and we lack high angular- and spectral-resolution surveys in the appropriate tracers to constrain transition mechanisms.

[C\,{\sc ii}] emission is closely related to the transition of the gas in the ISM between diffuse and dense phases. The [C\,{\sc ii}] 158 $\mu$m { line} is one of the brightest and widely distributed in the Milky Way, emitting up to 5\% of the total far-infrared (FIR) in photodissociation regions (PDRs), and functions as a coolant for the cold neutral medium \citep[e.g.][]{hollenbach97}. C$^+$ traces { regions} where H$^+$ is making the transition to H and H$_2$, since its ionization energy (11.6 eV) is lower than that of hydrogen. [C\,{\sc ii}] emission is found in HII regions, HI regions and in H$_2$ regions where the CO is photodissociated to C and C$^+$ \citep[e.g.,][]{pineda13, langer14, beuther14, velusamy15, pabst17}. Notably, [C\,{\sc ii}] is a tracer of the ``CO-dark molecular gas" component of the ISM that can { not} be seen in HI or CO line emission. It can directly distinguish HI clouds from diffuse inter-cloud HI gas, probe dense H$_2$ gas not associated with CO emission \citep{langer14}, and trace mass flows from CO cloud surfaces \citep{orr14}. 

The Carina Nebula Complex (CNC) is one of the most active star-forming regions in our galaxy. The CNC is roughly 400 times more luminous at optical wavelengths and 20 times larger in size than the Orion Nebula \citep[][and reference therein]{dias02,odell03}. This makes the CNC a prominent laboratory for studying the life cycle of the ISM { undergoing} extreme star formation. The CNC is also frequently compared to 30 Doradus, which is an extreme star-forming region in the Large Magellanic Cloud. The CNC harbors many massive stars (at least 70 O-type and WR stars, \citealt{smith06}) and has multiple phases of the ISM coexisting and transitioning from one to another as result of the strong radiation from the massive stars. Many observations have been carried out in emission lines and continuum bands to probe the structures in the nebula \citep[e.g.,][]{zhang01,brooks03,oberst11,preibisch12,young13,hartigan15,rebolledo16,rebolledo17,haikala17}. Integrated intensity maps in HI, H$\alpha$, [O\,{\sc i}], [C\,{\sc ii}], [C\,{\sc i}], CO, and dust continuum have revealed HII regions, PDRs, globules, and dense clouds \citep{brooks03,oberst11,hartigan15}. High-spectral resolution surveys in CO isotopologues and HI show a full complex of the diffuse and dense clouds in the nebula \citep{rebolledo16,rebolledo17}. Harboring various phases and structures of the ISM interacting with star formation, the Carina Nebula is a unique testbed to study the transition of the ISM associated with massive star formation. 

The ISM structure in the CNC has not been adequately probed due to lack of high spatial and spectral resolution observations in tracers such as [C\,{\sc ii}], [N\,{\sc ii}] and [O\,{\sc i}]. \citet{oberst11} carried out observations of [C\,{\sc ii}], [N\,{\sc ii}], and [O\,{\sc i}] using the South Pole Imaging Fabry-Perot Interferometer (SPIFI) and the Infrared Space Observatory (ISO). However, their study is limited to integrated intensity maps. The Mopra Southern Galactic Plane CO survey observed the CNC with a high spectral resolution of 0.088 km/s and revealed that molecular clouds have highly complex structures \citep{rebolledo16}. The CO spectral maps indicate that there are not only many molecular clouds and globules distributed throughout this region, but that there are also multiple velocity components along certain lines of sight within molecular clouds (e.g., Carina I \& II). The complexity of the structure of the molecular gas indicates that the ISM structures in the CNC must be probed using observations with high spatial and spectral resolution. 

In this study, we report a high spatial and spectral resolution survey toward the Trumpler 14 and Carina I (Tr 14/Carina I) region in the [C\,{\sc ii}] 158 $\mu$m transition using the Stratospheric Terahertz Observatory 2 (STO2). Tr 14 is an open cluster located half a degree (20 pc) west of the blue variable star $\eta$ Carinae, and Carina I is a dense cloud forming a dust lane located to the south of Tr 14, and illuminated by both Tr 16 (located 30' to the East of Tr 14 and containing $\eta$ Carinae) and Tr 14. The Tr 14/Carina I region contains multiple phases of the ISM in HII regions, PDRs, and dense molecular clouds. The key output of our survey is a high spatial and spectral resolution data cube in the [C\,{\sc ii}] 158 $\mu$m transition towards the Carina Nebula. { Here, we use} our [C\,{\sc ii}] map to study { physical structures of the ISM} including PDRs, molecular clouds, and HII regions in the CNC.

{ This study contains extensive analysis with many new findings. The followings are the highlights of this study and can be found in the discussion and conclusions sections. Comparing our [C\,{\sc ii}] spectral map to the Mopra CO map, we found that bright [C\,{\sc ii}] emission in the CNC is closely related to the CO clumps in position-position-velocity space, suggesting that bright [C\,{\sc ii}] emission likely arises from PDR and ionization fronts. We also found large absorption cavities in HI 21 cm emission and that those cavities are in a good agreement with the CO clouds/clumps and the Keyhole Nebula, a CO-dark molecular cloud, in position-position space. On the other hand, velocities of the absorption cavities are $\pm$5-10 km s$^{-1}$ shifted from the CO velocity centroids, suggesting that the cavities may follow cold HI gas photoevaporating or stripped from cloud surfaces. Through detailed PDR modeling of 10 different regions representing various ISM structures, we found a mass proportion of 1:4.1:5.6 for the atomic:dark:molecular(CO) gas and that six out of ten regions are dominated by [C\,{\sc ii}] emission from HII regions rather than PDRs. Finally, combining kinematics and modelings, we found that the three-dimensional morphology of the CNC is consistent with one side of numerous blister HII regions expanding freely toward us, similar to a champagne flow, with a lifetime of CO clouds exposed to HII regions being 20--30 Myr. }

We describe details of observations using STO2 and data reduction in \S\ref{sec:obs}. We show results and analysis along with complementary observations including dust continuum, CO, H92$\alpha$, HI 21cm, and Gaia Sky survey in \S\ref{sec:results:int} and \S\ref{sec:results:kin}. In \S\ref{sec:results:pdr}, we present detailed modeling of PDRs and its implications for the ISM structures in the Tr 14/Carina I region. In \S\ref{sec:dis}, we discuss a possible three-dimensional morphology of the Tr 14/Carina I region and uncertainties of data. We also discuss photoevaporation and mass loss of GMCs by EUV. Finally, in \S\ref{sec:con} we summarize our results.

\section{OBSERVATION \& DATA REDUCTION} \label{sec:obs}

Stratospheric Terahertz Observatory 2 is a balloon-borne observatory designed to fly in the stratosphere at 38 km altitude to avoid the severe atmospheric absorption at submillimeter wavelengths from ground-based sites. STO2 consists of a 0.8m telescope, a terahertz heterodyne receiver, and a high-resolution FFT spectrometer { (1 MHz)} with 1024 channels. STO2 was launched on December 7, 2016 and flew until Dec 29, 2016 over the Antarctic continent { and surveyed galactic plane and star-forming regions including the CNC.} The observations were made in two modes: On-The-Fly (OTF) mapping and spiral mapping. The OTF observations were done with a typical spacing of a half beam size { between raster observation lines} and relatively short integration time (0.65 seconds) per OTF dump, while the spiral observations were made with a sparse pointing ($>$2 FWHM beam size) and longer integration { times} ($>$1 second). { The maximum observation duration per raster line is set to be smaller than 35 seconds, which is the typical Allan variance time of the STO2 receivers. The telescope pointing was controlled by an on-board star tracker, and the typical pointing accuracy during the OTF mode was measured to be less than 15$''$.} For more details about the STO2 instrument and mission see Walker et al. (in prep). 

\begin{deluxetable}{cc}
\tablecolumns{2}
\tablewidth{0pt}
\tablecaption{Telescope properties\label{tab:prop}}
\startdata
\\
Facility & STO2  \\
Primary Diameter & 80 cm \\
Rest Frequency & 1900.537 GHz \\
Beam Size & 48$''$\\
Effective Resolution$^a$ & 55$''$\\
Pointing Accuracy & $<$15$''$\\
Spectral Resolution & 1 MHz\\
Velocity Resolution & 0.17 km s$^{-1}$\\
\enddata
\tablenotetext{a}{Angular resolution of the regridded map.}
\end{deluxetable}

Observations of the CNC were centered { on a position near the center of the Tr 14 cluster} at ($l$,$b$) = +287.33, -0.601 covering 0.25$^\circ$ by 0.28$^\circ$ in Galactic coordinates. { The observations were done in the OTF mode.} One OTF scan contains 45--47 spectra ({ $\sim$12$''$ spacing}) observing a 0.14$^\circ$ strip in Galactic latitude, which is half the map size. The native beam size at 158 $\mu$m is 48$''$ (0.53 pc at a distance of 2.3 kpc). { The distance to the CNC is still debated from reported between 2.2 kpc and 2.8 kpc, see \citealt{smith06,smith08,gaia16a,gaia16b,lindegren16,astraatmadja16,smith17} for more details.}. Spectra cover { an LST velocity range} from -112 km s$^{-1}$ to 57 km s$^{-1}$ { with a spectral resolution of 1 MHz (0.17 km s$^{-1}$ at 1.9 THz)}. For subtraction of broadband emission, observations toward a nearby reference position were made at the beginning and the end of each OTF scan. { We selected a reference position ($l$,$b$ = +286.50, +0.200)} based on the lowest [C\,{\sc i}] $^3P_1$--$^3P_0$/CO $J$ = 4--3 intensity ratio, indicative of minimal [C\,{\sc ii}] emission \citep{zhang01}. { The spectra toward the reference position are typically of good quality but vary slowly in time. To obtain an accurate reference spectrum for a given OTF scan, we linearly interpolated the reference spectra in time.} The single sideband system temperature was { 3,300 K with a typical variation of 100 K at 1.9 THz during the CNC observations.}

The STO2 data were reduced using the STO2 pipeline (Seo et al. in prep.). The STO2 pipeline is designed to process spectral scans considering unique characteristics found in the STO2 data. For example, some of the STO2 spectra have large fringes { ($>$50 K)} with their patterns varying over short periods { ($<$ 60 seconds)}. We could not effectively defringe the data with conventional observation software (e.g., CLASS). We wrote the STO2 pipeline to suppress large fringes by interpolating reference scans and using machine learning algorithms. { The machine learning algorithms characterized large-amplitude fringe patterns (using e.g., deflation independent component analysis, \citealt{hyvarinen00}) and identified extremely noisy spectra (using clustering algorithms on spectrum properties)}. { For the CNC observations, 90\% of the spectra were of sufficient quality to be included in the final spectral map.} In the reduced spectra, the typical overall noise level is 1.3 K in main-beam temperature, which is slightly larger than the expected radiometric noise (0.8 K) due to { the residual effects of fringe}. { The re-gridding of spectra was done following algorithm shown in \citet{mangum00} using a Gaussian-Bessel kernel without any weighting but omitting out exceptionally noisy spectra. The final effective beam size in the spectral map is 55$''$.} 

Intensities of the STO2 observations toward the CNC were calibrated using ISO [C\,{\sc ii}] data \citep{oberst11}. There were 12 positions observed by both STO2 and ISO. { From  a comparison of observed intensities toward these positions we estimate the main beam efficiency of STO2 to be 0.7$^{+0.14}_{-0.08}$}. { We therefore adopt a main beam efficiency of { 0.7} in the analysis of STO2 data (for more detail, see Appendix \ref{app:cal})}.

\begin{figure*}
\centering
\includegraphics[angle=0,scale=0.32]{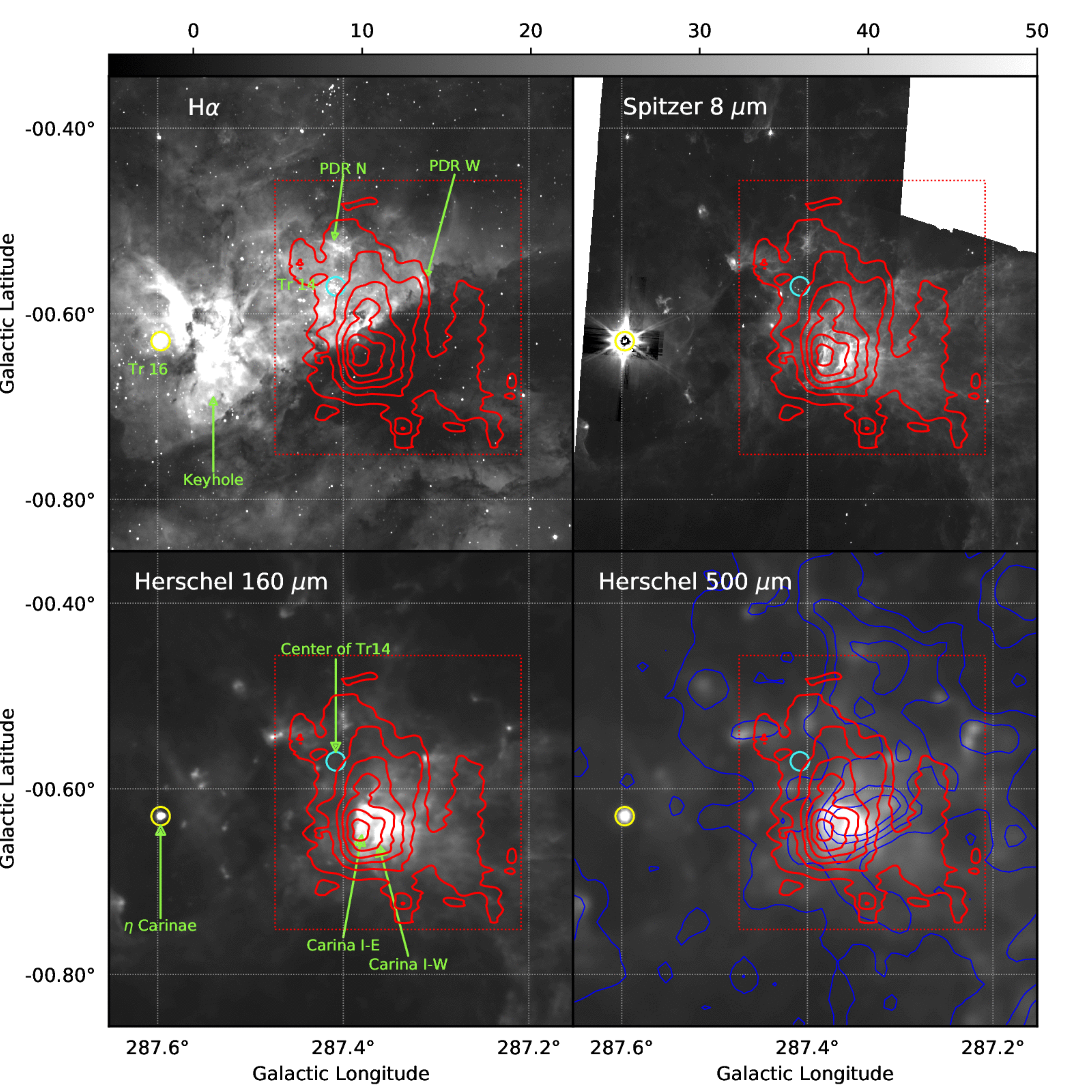}
\caption{Integrated intensity of [C\,{\sc ii}] 158 $\mu$m (red contours) overlaid on H$\alpha$, 8 $\mu$m, 160 $\mu$m, and 500 $\mu$m continuum images. The images are displayed in Galactic coordinates. The [C\,{\sc ii}] intensity contours are at 100, 150, 200, 250, 300, and 350 K km s$^{-1}$. The [C\,{\sc ii}] emission is integrated from -40 km s$^{-1}$ to 0 km s$^{-1}$. The box defined by the red dotted lines indicates the area mapped in [C\,{\sc ii}]. The blue contours in the bottom-right panel are the integrated intensity of $^{12}$CO 1--0 at levels of 20, 60, 100, 140, and 180 K km s$^{-1}$. The $^{12}$CO 1--0 emission is integrated from -40 km s$^{-1}$ to 0 km s$^{-1}$. The grayscale bar gives the intensity of the continuum images; the units for the H$\alpha$, 8 $\mu$m, 160 $\mu$m, and 500 $\mu$m images are 400 counts \citep[see][]{smith06}, 200 MJy/sr, 400 MJy/sr, and 200 MJy/sr respectively. The yellow and cyan circles denote $\eta$ Carinae, which is a member of Tr 16, and the center of Tr 14. The Tr 14 and Tr 16 clusters are also indicated in the first panel. The bright PDRs, the Keyhole nebula, Carina I-E, and Carina I-W are indicated by the green arrows. }
\label{fig_int_continuum}
\end{figure*}

\begin{figure*}
\centering
\includegraphics[angle=0,scale=0.31]{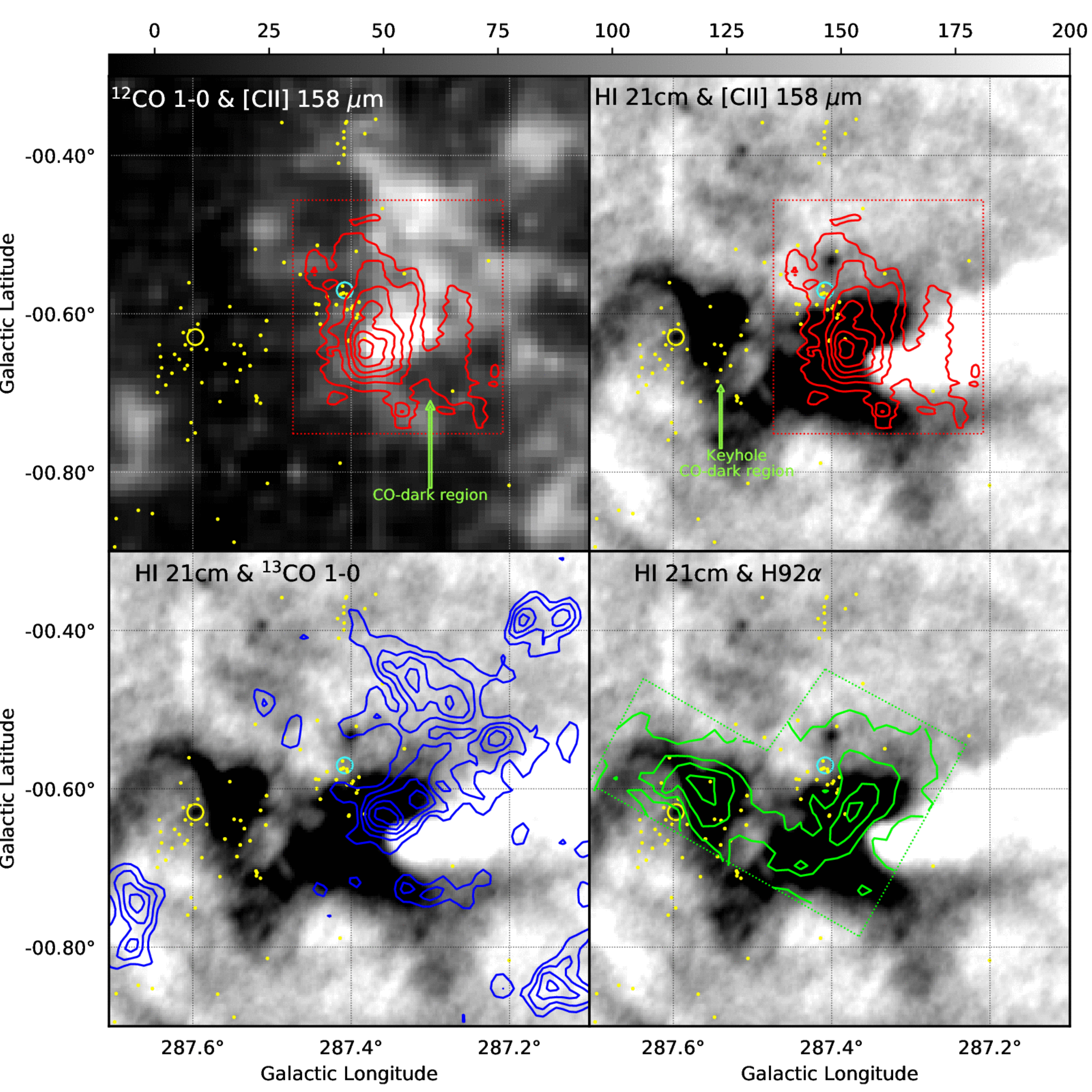}
\caption{Integrated intensity of [C\,{\sc ii}] 158 $\mu$m (red contours) overlaid on the integrated $^{12}$CO and HI 21cm images. [C\,{\sc ii}], $^{12}$CO, and HI 21cm emission is integrated from -40 km s$^{-1}$ to 0 km s$^{-1}$. The images are displayed in Galactic coordinates. The [C\,{\sc ii}] intensity contours are at 100, 150, 200, 250, 300, and 350 K km s$^{-1}$. The [C\,{\sc ii}] emission is integrated from -40 km s$^{-1}$ to 0 km s$^{-1}$. The integrated intensity of $^{13}$CO 1--0 (blue) and H92$\alpha$ (green) emission are overlaid on the HI 21cm emission in the third and fourth panels. The $^{13}$CO 1--0 contours are at 8, 12, 16, 20, 24 K km s$^{-1}$ and the H92$\alpha$ contours are at 20, 40, 60, and 80 K km s$^{-1}$. The $^{13}$CO 1--0 emission is integrated from -40 km s$^{-1}$ to 0 km s$^{-1}$, and the H92$\alpha$ emission is integrated from -65 km s$^{-1}$ to 20 km s$^{-1}$. The red and green dotted boxes define the area of the [C\,{\sc ii}] and H92$\alpha$ observations, respectively. The grayscale bar shows the scale of the integrated $^{12}$CO and HI 21cm images. The scale values are to be multiplied by 0.8 K km s$^{-1}$ and 10 K km s$^{-1}$ for the integrated $^{12}$CO and HI 21cm images are 0.8 K km s$^{-1}$ and 10 K km s$^{-1}$, respectively. The yellow and cyan circles denote $\eta$ Carinae and the center of Tr 14. The yellow dots denote O- and B-type stars \citep{alexander17}. The CO-dark region in the Tr 14/Carina I region and the Keyhole nebula, which is another CO-dark cloud, are indicated by the green arrows. }
\label{fig_int_lines}
\end{figure*}

\section{INTEGRATED INTENSITY OF [C\,{\sc ii}] EMISSION} \label{sec:results:int}

\subsection{Spatial Distribution of Integrated [C\,{\sc ii}] Emission}\label{sec:results:int:global}

We present the integrated intensity of the [C\,{\sc ii}] emission in Figures \ref{fig_int_continuum} and \ref{fig_int_lines} together with the H$\alpha$ image from {\it Hubble} \citep{smith06}, the 8 $\mu$m image from {\it Spitzer} \citep{smith10,povich11}, the 160 $\mu$m \& 500 $\mu$m images from {\it Herschel} \citep{preibisch12,gaczkowski13,roccatagliata13}, the integrated HI 21 cm image from ATCA \citep{rebolledo17}, the integrated CO 1--0 emission from Mopra \citep{rebolledo16}, and the integrated H92$\alpha$ emission from DSN \citep{horiuchi12}. { Using multiple continuum and spectral line images}, we describe here structures of the Tr 14/Carina I region and investigate spatial distribution of the [C\,{\sc ii}] emission with respect to different ISM phases. 

{ The overall morphology of the Tr 14 and Tr 16 region is as follows:} the Tr 14/Carina I region is located at the western part in the CNC while the Tr 16 region is located in the eastern part. The Tr 14 cluster is partially surrounded by dense clouds including the dark ``V"-shaped dust lane (see H$\alpha$ in Figure \ref{fig_int_continuum}) in the south of Tr 14 (a.k.a Carina I) and dense CO clouds in the west and north of Tr 14 (see $^{13}$CO 1--0 contours in Figure \ref{fig_int_lines}). The east side of the Tr 14/Carina I region is open to the $\eta$ Carinae and Tr 16 region, but there is another dust lane to the east of $\eta$ Carinae, suggesting that the dust lane and dense clouds partially surround the $\eta$ Carinae and Tr 14 region. The optical image shows that a majority of the members of the Tr 14 and Tr 16 clusters have low extinction \cite[e.g.][]{smith06}, which indicates that there is no significant foreground cold gas toward the two clusters and the HII regions are exposed to us. On the other hand, the ``V"-shaped dust-lane appears as a high extinction region in the H$\alpha$ image, suggesting that the dust lane is in front of the HII region \citep[e.g.,][]{wu18}.



We find that the [C\,{\sc ii}] emission covers a significant fraction of the area mapped using STO2 (0.25$^\circ$ x 0.28$^\circ$ equivalent to 10.0 pc x 11.2 pc at a distance of 2.3 kpc). The fraction of the area with peak main beam temperature $>$5 K and $>$10K are 83\% and 58\%, respectively. [C\,{\sc ii}] emission is expected from both the HII region of Tr 14 and the PDRs. Looking at the H$\alpha$ and 8 $\mu$m maps, we find that the upper half of our [C\,{\sc ii}] map coincides with the HII region of Tr 14 and the other half of the [C\,{\sc ii}] map is coincident with the Carina I cloud and its PDRs, which confirms that there are multiple sources for the [C\,{\sc ii}] emission. The brightest intensity peak of the [C\,{\sc ii}] emission is 370 K km s$^{-1}$, located 7$'$ south of Tr 14 (4.7 pc at the distance of 2.3 kpc), where the Carina I-E/Carina I-W clouds are located (indicated by green arrows in Figure \ref{fig_int_continuum}). The brightest emission at Carina I-E/Carina I-W is because they are the densest clouds in Carina I and irradiated by a B1 supergiant only 0.5 pc away in projected distance, in addition to main members of the Tr 14 cluster, thus, resulting in significantly high emission measure (excitation condition of these clouds is further discussed in \S\ref{sec:results:pdr}).  

We compare the integrated [C\,{\sc ii}] emission to the dust continuum emission and PAH observed using {\it Herschel} and {\it Spitzer}. The dust continuum emission at 160 and 500 $\mu$m shows warm and cold dust structures in the Tr 14/Carina I region. The 8 $\mu$m emission from {\it Spitzer} is typically dominated by PAH emission, which traces PDRs in star-forming regions. Overall, the strong [C\,{\sc ii}] emission agrees better on a large scale with the bright structures seen in dust and PAH emission than with the bright structure of the H$\alpha$ emission, suggesting that the strong [C\,{\sc ii}] emission may originate from PDR and HII regions near ionization fronts, while we still observe weak [C\,{\sc ii}] emission coming from the inner part of the Tr 14 HII region.

{ We show the integrated $^{12}$CO and $^{13}$CO 1--0 observed using Mopra} in Figures \ref{fig_int_continuum} and \ref{fig_int_lines} \citep{rebolledo16} along with the integrated [C\,{\sc ii}] emission. The $^{12}$CO 1--0 map reveals the cold molecular ISM, and the $^{13}$CO 1--0 map highlights the denser portions of the CO clouds in the Tr 14/Carina I region. The overall spatial distribution of the CO 1--0 emission, particularly $^{13}$CO 1--0, shows that CO clouds form a wall bounding the western part of the Tr 14/Carina I region. There is also weak, broad $^{12}$CO 1--0 emission from Tr 14, suggesting that there may be CO gas behind Tr 14 since we do not see significant extinction in optical bands. We find that, overall, the [C\,{\sc ii}] emission is broadly distributed covering Tr 14 and nearby CO clumps, while the CO emission is bright 4$'$ west and 5$'$ south of Tr 14 (2.6 pc and 3.4 pc at a distance of 2.3 kpc) and extended to the western part of the Tr 14/Carina I region. There is a region with relatively bright [C\,{\sc ii}] emission (121 K km s$^{-1}$) but quite weak CO emission (integrated intensity 34 K km s$^{-1}$ compared to 211 K km s$^{-1}$ from Carina I-E), which may indicate a CO-dark molecular region \citep{langer14} (indicated by a green arrow at the first panel in Figure \ref{fig_int_lines}). The intensity peaks of individual CO clumps are typically displaced a couple of arcminutes relative to the [C\,{\sc ii}] and PAH (8 $\mu$m) intensity peaks, showing the locations of CO clumps relative to their PDRs and ionization fronts.


{ We probe spatial distribution of the cold neutral medium using the integrated HI emission and the [C\,{\sc ii}] emission (Figure \ref{fig_int_lines}).} We integrate the HI emission from -40 km s$^{-1}$ to 0 km s$^{-1}$ because most of the CO and [C\,{\sc ii}] emission is within this velocity range. The integrated HI 21cm emission shows extended HI emission covering the CNC with clumpy HI clouds, and with cavities in the HI emission near Tr 14 and $\eta$ Carinae. Cavities may appear when neutral atomic hydrogen forms molecular hydrogen, becomes ionized, or if there is foreground absorption due to cooler HI gas. The cavities in the CNC are due to the absorption features in the HI spectra. We also find that the spatial distribution of the cavities agrees well with locations of the Keyhole Nebula and the dense CO clouds near Tr 14. To have absorption features, there must be a continuum background. In the CNC, the free-free emission of the HII region provides a bright, hot background against which we see absorption. We find that the distribution of the H92$\alpha$ emission is broadly extended including the HII region and the cavities. Comparing HI 21cm to [C\,{\sc ii}] and $^{12}$CO 1--0, we find numerous bright CO clumps and bright [C\,{\sc ii}] emission within the cavities in the Tr 14/Carina I region. In this part of the cavity, the cold HI in dense CO clumps may be the source of absorption, which was also discussed in \citet{rebolledo17}. On the other hand, we do not see significant CO emission in the cavity west of $\eta$ Carinae, where the Keyhole nebula is located. This suggests that the Keyhole nebula is a CO-dark cloud with relatively cold HI together with HII gas.  


\section{KINEMATICS OF [C\,{\sc ii}] AND OTHER TRACERS IN THE TR 14/CARINA I REGION} \label{sec:results:kin}

\begin{figure*}
\centering
\includegraphics[angle=0,scale=0.32]{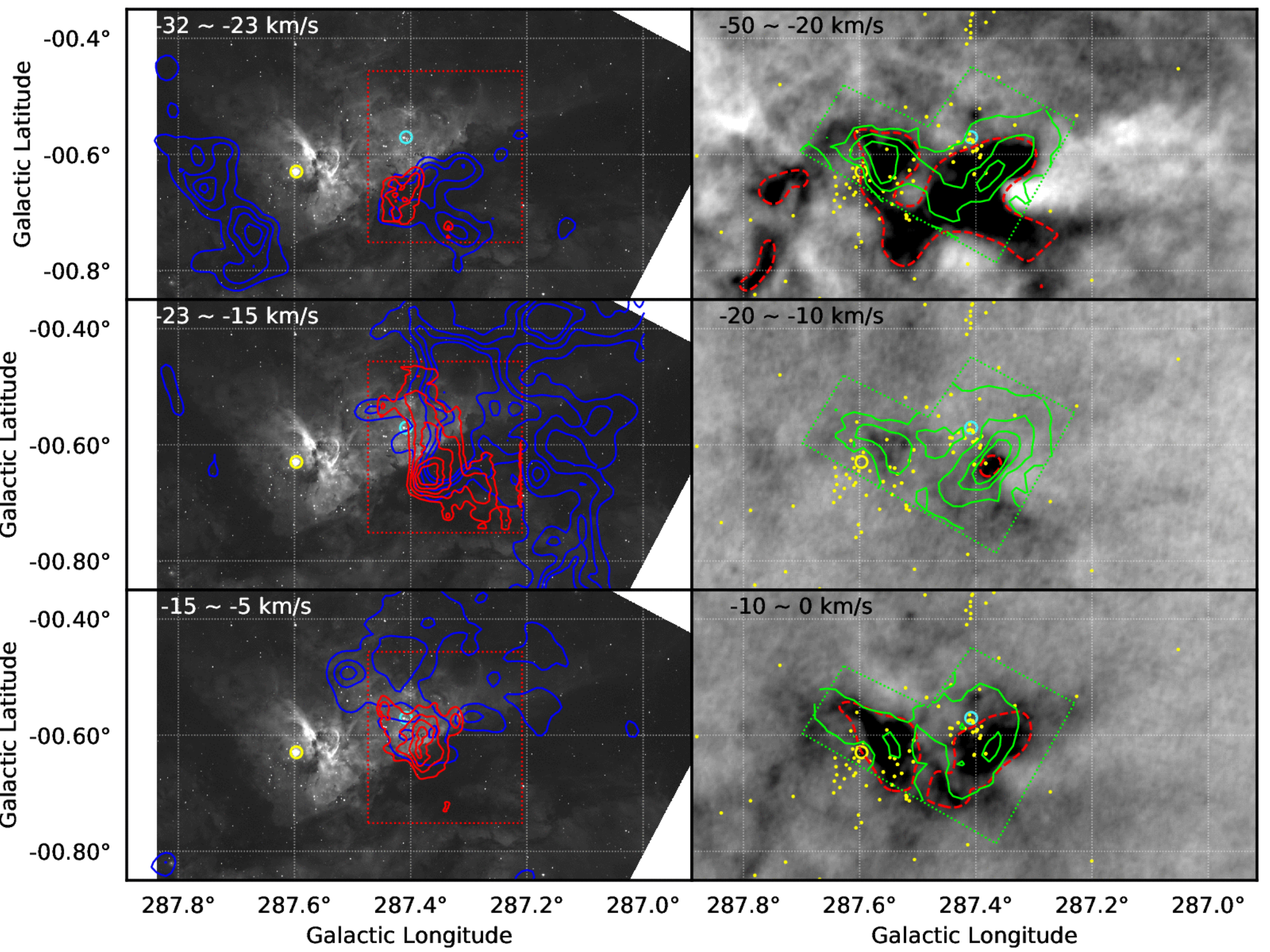}
\caption{{ Channel maps of [C\,{\sc ii}] 158 $\mu$m obtained with STO2 (red contours), $^{12}$CO observed with Mopra (blue contours, \citealt{rebolledo16}), H92$\alpha$ (green contours, \citep{horiuchi12}), and HI 21cm (grayscale background in the right column, \citealt{rebolledo17}). The grayscale background in the left column is an H$\alpha$ image \citep{smith06}. The [C\,{\sc ii}] contours start at 40 K km s$^{-1}$ and increase in 20 K km s$^{-1}$ increments. The $^{12}$CO contours start at 10 K km s$^{-1}$ and increase in 10 K km s$^{-1}$ increments. The H92$\alpha$ contours are at 10, 20, 30 and 40  K km s$^{-1}$ in the top panel and are at 5, 10, 15, and 20 K km s$^{-1}$ in the middle and bottom panels. The red and green boxes with the dotted lines are the areas mapped in [C\,{\sc ii}] and H92$\alpha$, respectively. The red dashed contours are at an antenna temperature of -50  K km s$^{-1}$ in HI 21 cm to show the absorption cavities. The large yellow and cyan circles indicate $\eta$ Carinae and the center of the Trumpler 14 cluster. The yellow dots denote O- and B-type stars \citep{alexander17}.}}
\label{fig_int_chan}
\end{figure*}

{ Here we} discuss the structure and kinematics of ionized, neutral, and molecular gas in the Tr 14/Carina I region {  in position-position-velocity (PPV) space}. We analyze the channel maps and spectra of our [C\,{\sc ii}] 158 $\mu$m observation along with other observations including H92$\alpha$ \citep{horiuchi12}, $^{12}$CO \& $^{13}$CO 1--0 \citep{rebolledo16}, H$\alpha$ \citep{smith06}, and HI 21 cm \citep{rebolledo17}. To disentangle the complicated ISM structure in the CNC, we first focus on the dense cloud/clumps and their PDRs traced by CO and [C\,{\sc ii}], and expand our view to the ionized and neutral atomic media traced by optical lines (H$\alpha$, [N\,{\sc ii}] 6548$\rm \AA$, \citealt{damiani16}), by radio recombination line (H92$\alpha$, \citealt{horiuchi12}), and by HI 21cm \citep{rebolledo17}.

\subsection{Channel Maps of [C\,{\sc ii}] 158 $\mu$m and CO 1--0} \label{sec:results:kin:CO}

Figure \ref{fig_int_chan} shows the channel maps of [C\,{\sc ii}] 158 $\mu$m, $^{12}$CO 1--0 \citep{rebolledo16} overlaid on H$\alpha$ \citep{smith06}. The [C\,{\sc ii}] emission is mostly found in the velocity range from -32 km s$^{-1}$ to -5 km s$^{-1}$ and is spatially localized near Tr 14. We find that the $^{12}$CO 1--0 emission also covers the velocity range from -32 km s$^{-1}$ to -5 km s$^{-1}$ but spans a wider area west of Tr 14 compared to the [C\,{\sc ii}] emission. We find no significant CO emission between $\eta$ Carinae and Tr 14 in any channel map. We find weak CO emission east of $\eta$ Carinae but at slightly blue-shifted velocity (-25 km s$^{-1}$) compared the CO emission west of Tr 14 (-20 km s$^{-1}$).

Based on the distribution of the [C\,{\sc ii}] 158 $\mu$m and $^{12}$CO 1--0 emission in PPV space, we may divide the dense structures into { three velocity groups.} The first group includes the $^{12}$CO and [C\,{\sc ii}] structures in the LSR velocity range from -32 km s$^{-1}$ to -23 km s$^{-1}$, the second group includes the dense structures in the velocity range from -23 km s$^{-1}$ to -13 km s$^{-1}$, and the third group includes the dense structures in the LSR velocity range from -13 km s$^{-1}$ to -5 km s$^{-1}$.

The first group (-32 to -23 km s$^{-1}$) contains a few CO clumps south of Tr 14 including a part of Carina I-E/Carina I-W and the CO clumps east of $\eta$ Carinae. This group is likely in front of Tr 14 relative to us since CO clumps in this velocity range appear as dark clumps with respect to the bright H$\alpha$ background \citep{haikala17}. Also, their blue-shifted velocity compared to the LSR velocity of most of the CO clumps suggests that they are pushed toward to us by expanding HII gas and in foreground { of the HII region}. The CO clumps in this group are relatively isolated from each other and have bright [C\,{\sc ii}] layers on their outskirts (for more detail, see the maps at -28.5 and -23.5 km s$^{-1}$ in Appendix \ref{app:channel}), which indicates the presence of { ionization fronts and} PDRs surrounding those CO clumps due to high-mass stars in the Tr 14/Carina I region. We find that the [C\,{\sc ii}] emission is mostly located in the eastern outskirts of the CO clumps rather than in the northern outskirts facing { the center of} Tr 14. This may be due to the B1 supergiant and O7 binary on the east side of Carina I \citep{wu18}. The { spatial distribution of CO and [C\,{\sc ii}] emission} suggests that the CO clumps in the first group may not be at the same distance from us as is Tr 14, which is consistent with silhouette globules at velocity range from -30 km s$^{-1}$ to -20 km s$^{-1}$ facing towards Tr 16 rather than Tr 14 \citep{smith03}.   


Near $\eta$ Carinae, we see that there are CO clumps to the east at -28.5 km s$^{-1}$, which comprises the east dust lane of the CNC. These CO clumps are blue-shifted compared to the CO cloud to the west and their LSR velocity is similar to that of the CO clumps in the dust lane south of Tr 14. This suggests that CO clumps to the east of $\eta$ Carinae are likely in the foreground of the high-mass stars in the CNC. We do not see any strong CO emission near $\eta$ Carinae in any channels, while there are CO clumps near Tr 14 in the red-shifted velocity range of -16 to -8 km s$^{-1}$. It is likely that $\eta$ Carinae may have cleared out the dense structures where it originally formed, while Tr 14 is still interacting with nearby dense clumps. This { picture} is consistent with the younger age of Tr 14 compared to Tr 16 \citep{walborn73,morrell88,vazquez96,smith08,rochau11}, suggesting that Tr 14 has not lived long enough to clear out its surroundings. We will further discuss the three-dimensional structure of the Tr 14/ Carina I region in \S\ref{sec:dis}.

The second group (-23 to -15 km s$^{-1}$) includes the CO clouds/clumps west of Tr 14. In this velocity range, we find that the CO emission is brighter and more extended than the [C\,{\sc ii}] emission in the other groups, suggesting the majority of the CO gas is within this velocity range. The spatial distribution shows highly clumpy CO structures but also includes a dense CO ``wall" to the west of Tr 14 in the LSR velocity range from -20 km s$^{-1}$ to -15 km s$^{-1}$ { with their central velocity near -17 km s$^{-1}$.} The [C\,{\sc ii}] emission reveals isolated structures associated with the CO clumps in the LSR velocity range from -23 km s$^{-1}$ to -20 km s$^{-1}$. The [C\,{\sc ii}] emission associated with the CO clump is likely due to the PDRs of the clump. On the other hand, at velocities from -20 to -15 km s$^{-1}$, we find that the [C\,{\sc ii}] emission forms a thick strip following the eastern outskirts of the CO wall. The bright [C\,{\sc ii}] strip is the ionization front of the dense CO wall.    

In the third group (-15 to -5 km s$^{-1}$), the $^{12}$CO emission is found on and around the center of Tr 14. Considering that the extinction from H$\alpha$ is quite low in this direction \citep{smith06,hur15}, we think that the CO gas in this group is likely behind Tr 14 along our line of sight. In the [C\,{\sc ii}] channel maps, we find the majority of the [C\,{\sc ii}] emission is found to be spatially associated with CO clumps. { (e.g., see channel map at -11 km s$^{-1}$ in Appendix \ref{app:channel}).} This indicates that the [C\,{\sc ii}] emission in this group comes from the PDRs of the CO clumps, which are behind Tr 14 and may be being pushed away from us.

Beyond -5 km s$^{-1}$, there is weak $^{12}$CO 1--0 emission around Tr 14 but we could not find a significant dense cloud. This suggests that the dense structures of the Tr 14/Carina I are mostly within the -32 -- -5 km s$^{-1}$ velocity range.

\subsection{Channel maps of [C\,{\sc ii}] 158 $\mu$m, CO 1--0, HI 21cm, and H92$\alpha$} \label{sec:results:kin:others}

We compare the [C\,{\sc ii}] 158 $\mu$m and $^{12}$CO 1--0 emission to the H92$\alpha$ emission in order to investigate the distribution of the ionized ISM in the Tr 14/Carina I region { (right columns of Figures \ref{fig_int_chan} and channel maps in Appendix \ref{app:channel}).} We find that ionized gas is distributed slightly asymmetrically between Tr 16 and Tr 14 in velocity space. The H92$\alpha$ emission near Tr 16 and $\eta$ Carinae spans -60 km s$^{-1}$ to +15 km s$^{-1}$, while the H92$\alpha$ emission near Tr 14 is found from -40 km s$^{-1}$ to +10 km s$^{-1}$. This suggests that the HII regions near Tr 16 and Tr 14 have different dynamics or spatial distributions. The spatial distribution of the H92$\alpha$ emission has intensity peaks at two different locations: one is the Keyhole Nebula and the other is Carina I-E, which are both dense clouds near $\eta$ Carinae and Tr 14. We see the brightest intensities toward the dense clouds rather than toward { the inner HII region of Tr 14} because the column densities of both the electrons and hydrogen atoms near the dense clouds are higher. The H92$\alpha$ emission is extended in the CNC, indicating that the ionized gas is widely distributed.

We first compare the HI 21cm emission to the H92$\alpha$, [C\,{\sc ii}] 158 $\mu$m, and $^{12}$CO 1--0 emission to probe the distribution of the neutral medium to ionized and molecular media. The large-scale structure of HI in the CNC is discussed in \citet{rebolledo17}, so we focus on the small-scale structures ($<$ 20 pc) of the HI emission within the Tr 14/Carina I region. 


We find that there are cavities in the HI channel maps due to absorption features in the HI spectra as shown in \citet{rebolledo17}. Comparing the cavity to the [C\,{\sc ii}] and CO emission, we find that the cavities have a strong correlation with CO and [C\,{\sc ii}] in PPV space. In position space, we see that the west portion of the cavity coincides with Carina I-E, while the east portion of the cavity coincides with the Keyhole Nebula. In velocity space, the HI cavities are at two different velocity ranges: one at -50 to -20 km s$^{-1}$ and the other at -10 to 0 km s$^{-1}$. The CO gas is at LSR velocities between the two velocity ranges of the HI cavities (-30 to -5 km s$^{-1}$). These observations indicate that the neutral atomic medium is spatially associated with the cold molecular clouds but has different kinematics with respect to the dense molecular gas (e.g., cloud dispersal through stripping and photoevaporation). 

\subsection{Spectra of CO 1--0, [C\,{\sc ii}] 158 $\mu$m, HI 21cm, H92$\alpha$, and Optical Lines} \label{sec:results:kin:spectra}

\begin{figure*}
\centering
\includegraphics[angle=0,scale=0.31]{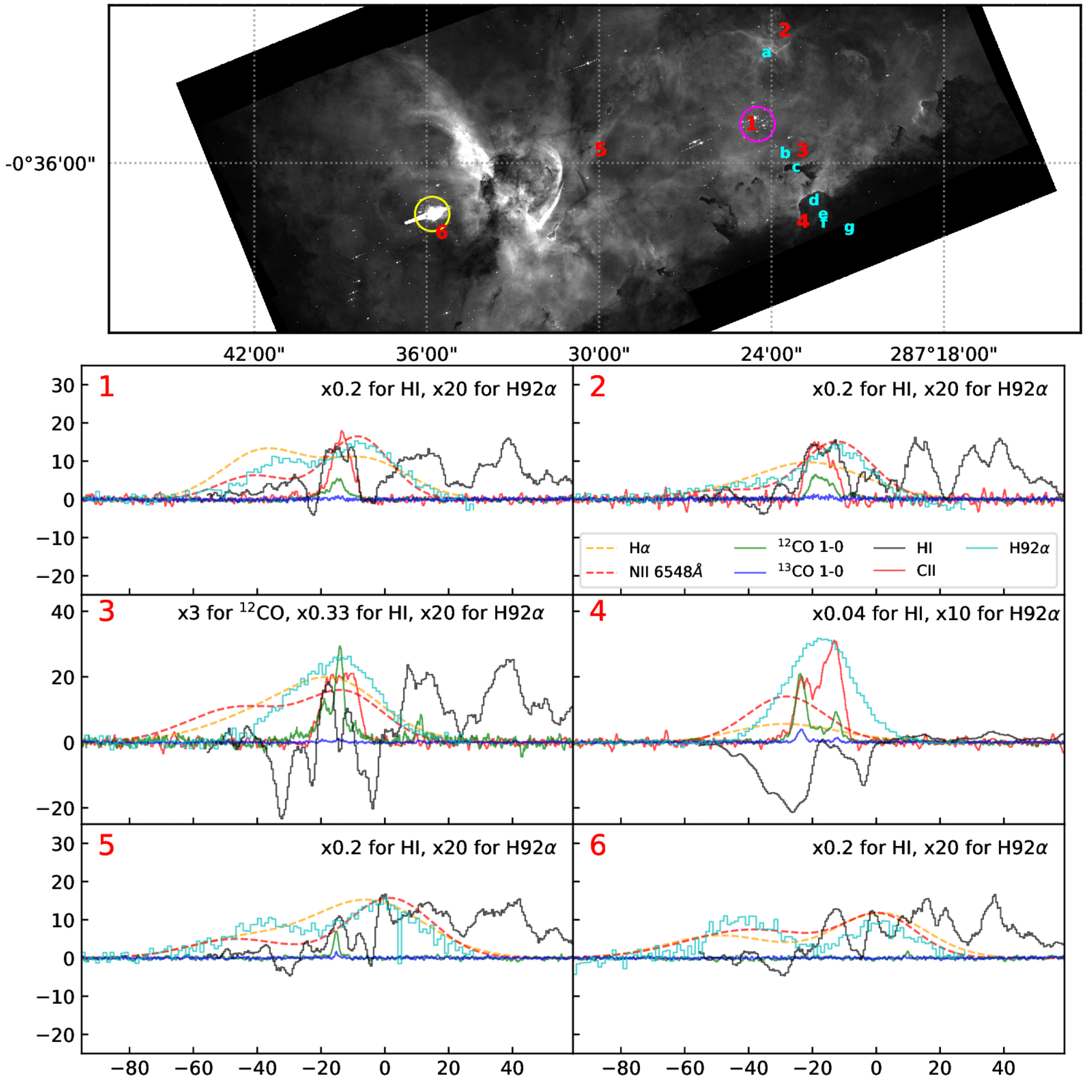}
\caption{Top panel: 6 positions employed for comparison of spectra indicated by numbers and 7 positions adopted for PDR modeling indicated by alphabet letters on the H$\alpha$ image. The yellow and magenta circles denote $\eta$ Carinae and Tr 14. Bottom panels: Spectra of [C\,{\sc ii}] 158 $\mu$m, $^{12}$CO and $^{13}$CO 1--0 \citep{rebolledo16}, HI 21cm \citep{rebolledo17}, H92$\alpha$ \citep{horiuchi12}, H$\alpha$ and [N\,{\sc ii}] 6548$\rm \AA$ \citep{damiani16} toward 6 selected positions in the Tr 14/Carina I region. The x-axis is the LSR velocity in km s$^{-1}$. The y-axis is the main beam temperature in Kelvin for [C\,{\sc ii}] 158 $\mu$m, $^{12}$CO 1--0, and $^{13}$CO 1--0. For HI 21 cm and H92$\alpha$, the intensities are in units of Kelvin but with a scaling factor applied to displayed spectra written in each panel. The y-axis is the relative intensity for the optical lines. The H$\alpha$ and [N\,{\sc ii}] 6545$\rm \AA$ are the best-fit models to the Gaia Sky spectra toward the CNC, while the other spectra are the observed spectra.}
\label{fig_lines}
\end{figure*}
\begin{figure*}
\centering
\includegraphics[angle=0,scale=0.31]{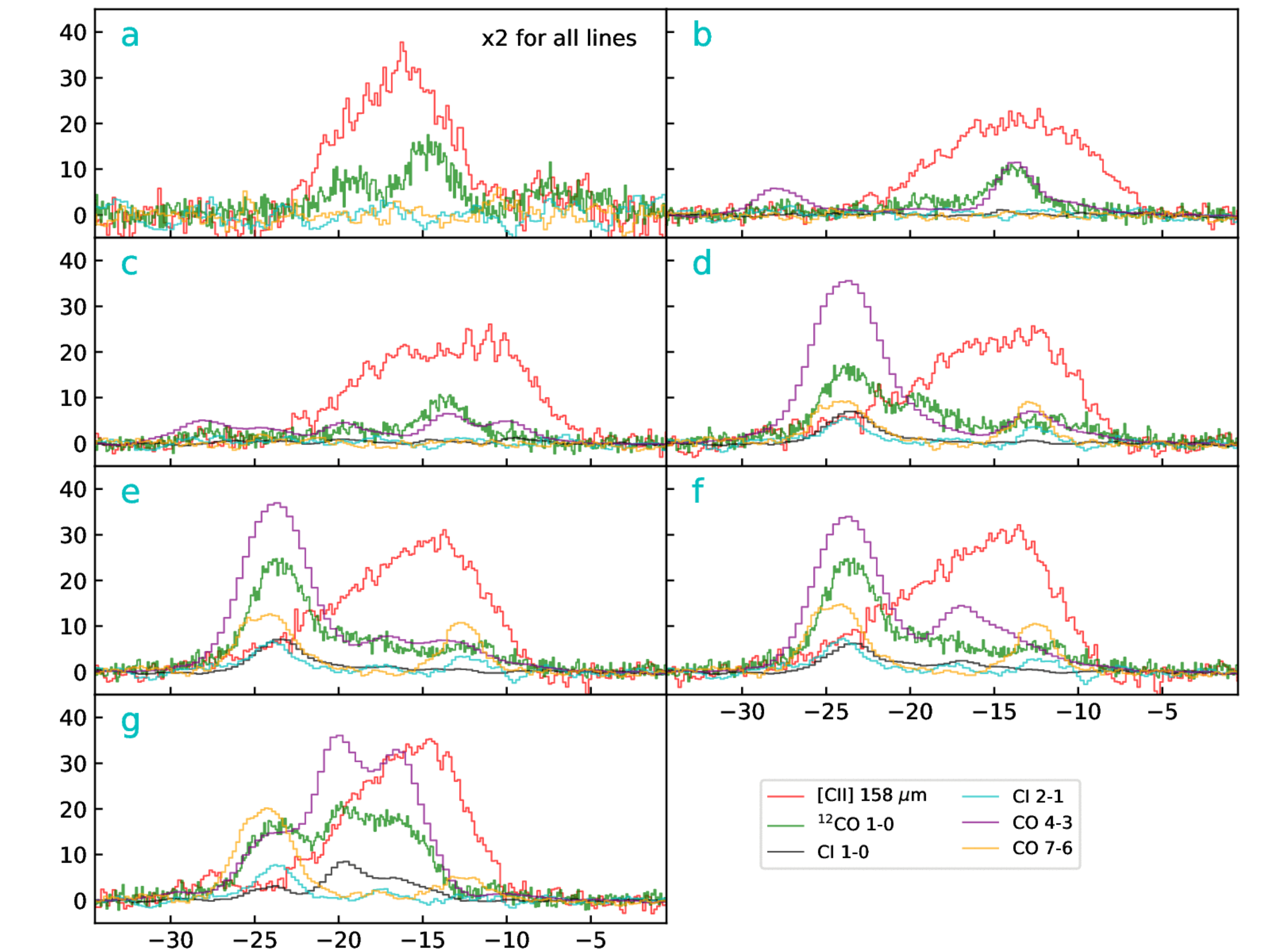}
\caption{Spectra continued from Figure \ref{fig_lines}. Spectra of [C\,{\sc ii}] 158 $\mu$m, [C\,{\sc i}] $3P_1$--$3P_0$ and $3P_2$--$3P_1$ \citep{kramer08}, and CO 4--3 and 7--6 \citep{kramer08} toward 7 selected positions for PDR modeling in the Tr 14/Carina I region. The x-axis is the LSR velocity in km s$^{-1}$. The y-axis is the main beam temperature in Kelvin.}
\label{fig_pdr_lines}
\end{figure*}

{ To study the ISM structure,} we select 6 positions representative of the HII region in the CNC, the ionization front of the large CO cloud to the west of Tr 14, and molecular regions (Figure \ref{fig_lines}, numbers 1--6). We analyze the spectra including $^{13}$CO 1--0, $^{12}$CO 1--0 \citep{rebolledo16}, [C\,{\sc ii}] 158 $\mu$m, HI 21cm \citep{rebolledo17}, H92$\alpha$ \citep{horiuchi12}, and optical lines of nitrogen and hydrogen \citep{damiani16}. 

Panels 1, 5 and 6 in Figure \ref{fig_lines} present the spectra towards the three positions including the center of Tr 14, the middle position between Tr 14 and $\eta$ Carinae, and $\eta$ Carinae respectively. We analyze these spectra to trace the kinematics of the HII region because Positions 1 and 6 are the centers of the HII regions formed by Tr 14 and Tr 16, respectively, and Position 5 is at the interface of the two HII regions. We find three common features among the spectra of H92$\alpha$, H$\alpha$, and [N\,{\sc ii}] 6548$\rm \AA$: wide velocity ranges compared to CO and [C\,{\sc ii}], double intensity peaks, and long tails toward negative velocities. The H92$\alpha$, H$\alpha$, and [N\,{\sc ii}] 6548$\rm \AA$ spectra at all three positions cover velocities from -80 km s$^{-1}$ to +30 km $^{-1}$. These velocity ranges are at least a factor of two larger than those in the other three positions representing PDRs and ionization fronts. 

We find double peaks in the H92$\alpha$, H$\alpha$, and [N\,{\sc ii}] 6548$\rm \AA$ profiles at positions 1, 5, and 6. The blue-shifted intensity peaks are at -40 to -30 km s$^{-1}$ and the red-shifted intensity peaks are at -5 to +5 km s$^{-1}$. As discussed in \citet{damiani16}, the double intensity peaks likely indicate the red- and blue-shifted boundaries/shells of the HII region, since { their} emission measure is expected to be the highest toward the ionization front due to photoevaporation \citep[e.g.,][]{krumholtz07}. We find that the line widths of the peaks are significantly larger ($>$30 km s$^{-1}$ for H$\alpha$, and $>$25 km s$^{-1}$for [N\,{\sc ii}] 6548$\rm \AA$) than the thermal broadening (21 km s$^{-1}$ for H$\alpha$, and 5.7 km s$^{-1}$ for [N\,{\sc ii}] 6548$\rm \AA$ at { at 10,000 K}), indicating the presence of considerable dynamical motions such as expansion of the HII regions and turbulence. We also find that the blue-shifted portions of the spectra ($<$-20 km s$^{-1}$) typically have long tails toward negative velocities. For example, in the spectra at position 6 ($\eta$ Carinae) we see that the blue-shifted intensity peak is at -40 km s$^{-1}$ and the profile extends to -80 km s$^{-1}$, while the red-shifted intensity peak is at 0 km s$^{-1}$ and the profile extends only to +20 km s$^{-1}$. We see similar profiles towards positions 1 and 5 but with smaller separations between the two intensity peaks and shorter tails compared to the spectra at position 6. The long-tails on the negative velocity side of the spectra indicate that the blue-shifted portion of the HII region is likely larger and expanding faster than the red-shifted portion. Considering that there is almost no blue-shifted HI 21cm emission and that most of the high-mass stars show low extinctions in Tr 14 and Tr 16, it appears that the blue-shifted portion of the HII region has burst through the dense gas and is freely expanding towards us while the red-shifted portion of the HII region is confined by an HI cloud, similar to the Champagne model \citep{tenorio-tagle79}.

Comparing the spectra towards positions 1, 5, and 6, we find that there is CO emission toward Positions 1 (Tr 14) and 5 (a middle position between Tr 14 and $\eta$ Carinae), while we do not find any significant CO emission toward position 6 ($\eta$ Carinae). This suggests that CO clumps in Tr 14 are still confined and are interacting with the HII region, while Tr 16 has mostly cleared out nearby dense structures except for the CO cloud east of $\eta$ Carinae, which is consistent with their ages \citep{walborn73,morrell88,vazquez96,smith08,rochau11}. We find that the CO emission at positions 1 and 5 is at a similar LSR velocity of -18 km s$^{-1}$, while the LSR velocities of the double-peaks in H92$\alpha$, H$\alpha$, and [N\,{\sc ii}] 6548$\rm \AA$ are considerably different towards the two positions. The trend is almost same throughout the entire Tr 14 region. This may indicate that the CO clouds at -18 km s$^{-1}$ are beyond the HII boundary without getting much acceleration by the expanding HII gas yet.

Position 2 shows the spectra toward a PDR in the north of Tr 14 (PDR N in Figure \ref{fig_int_continuum}). This region is considered to be behind Tr 14 since it shows bright emission on its entire surface in H$\alpha$ and 8 $\mu$m. We find $^{12}$CO and [C\,{\sc ii}] emission around -20 km s$^{-1}$, indicating that the CO clump has a PDR. We do not find any significant $^{13}$CO emission toward this position, indicating that the CO clump may not have high column density. We also see emission of H92$\alpha$, H$\alpha$, and [N\,{\sc ii}] 6548$\rm \AA$. However, the profiles of those lines show only a single intensity peak with a long tail toward negative velocities. We may not see the double peaks since this clump is close to the edge of HII region and the expanding HII gas motions are mostly tangential to our lines of sight. The long tail of the line profile toward to negative velocities may be due to the expansion of the HII region towards us.

Position 3 is towards an ionization front of the dense CO wall in the west of Tr 14. We find multiple components of $^{12}$CO emission along the line of sight and a single component of $^{13}$CO emission associated with the strongest $^{12}$CO component. We find relatively broad [C\,{\sc ii}] emission coinciding with the CO components around -20 km s$^{-1}$. The H92$\alpha$, H$\alpha$, and [N\,{\sc ii}] 6548$\rm \AA$ lines show different profiles but the velocities of their intensity peaks are around -18 km s$^{-1}$ which is near the velocity of the CO and [C\,{\sc ii}] intensity peaks. This is likely due to the high emission measure of the ionized gas near the PDRs of the CO clumps. We find skewed profiles towards negative velocities as similar as the spectra toward Tr 14, which is likely related to the HII region expanding toward us. 

Position 4 is towards Carina I-E, where we observed the brightest CO and [C\,{\sc ii}] emission. We find that there are at least two dense CO clumps along the line of sight. The CO component at -23 km s$^{-1}$ is likely in front of Tr 14 since we see it as a dark clump in contrast against a bright HI background but it also has bright [C\,{\sc ii}] emission, which indicates that there is a PDR. The entire surface of Carina I-E has significant H$\alpha$ and 8 $\mu m $ emission, which confirms that we see the PDR of Carina I-E largely face-on but from the back (non-illuminated) side. Another CO component is at -12 km s$^{-1}$. This component has significantly brighter [C\,{\sc ii}] emission with respect to the CO emission. The H92$\alpha$ line includes both of these velocities. H$\alpha$ and [N\,{\sc ii}] 6548$\rm \AA$ lines show single-intensity peaks and have their peaks at -25 km s$^{-1}$, which is different from double intensity peaks in other positions. This may be mainly due to high optical depth toward dense clumps and obscuring the other intensity peaks. These observations indicate that the component at -12 km s$^{-1}$ is an HII region interacting with a CO cloud. Comparing to positions 1, 5, and 6, we see that the H92$\alpha$, H$\alpha$, and [N\,{\sc ii}] 6548$\rm \AA$ lines at position 4 have narrower velocity ranges and have single intensity peaks while the line width ($\sim$27 km s$^{-1}$ for H92$\alpha$) is still significantly broader than the thermal line width (21 km s$^{-1}$ at 10,000 K). It is likely that the HII region is in between dense clouds (foreground and background) or it may be near the edge of the HII region where we would not see the expansion of the HII region along the line of sight.

The HI 21cm profiles towards positions 3 and 4 have complicated features including both emission and absorption. In the CNC, we find that there are two types of HI absorption features. One is HI absorption at the same velocity as $^{13}$CO emission. This is likely due to cold HI gas within the dense CO clumps, which may still have a relatively high HI column density due to high total column density in a CO clump combined with modest fractional abundance of HI due to relatively young age and incomplete conversion to H$_2$ \citep[][and reference therein]{wakelam17}. The absorption feature associated with $^{13}$CO intensity peak towards position 3 suggests that there are cold HI gas within the molecular clumps. The other absorption features are the ones that are red- or blue-shifted relative to the $^{12}$CO components. One possible explanation is evaporation or stripping from a CO clump by extreme radiation, since the photoevaporating or radiation stripped HI gas from the CO clumps can appear as absorption against the free-free emission background produced by the HII region. We find that the velocity differences between the absorption features and the CO components in the Tr 14/Carina I region are typically $\leq$10 km s$^{-1}$. These velocity differences are similar to the photoevaporation or radiation stripping velocity from the CO clouds predicted by numerical simulations \citep[e.g.,][]{bertoldi89,bertoldi90,lefloch94,mellema98,mcleod16}. In addition, the spatial distribution of the absorption features coincides with the dense clouds (e.g., Carina I-E/Carina I-W and Keyhole nebula) in the CNC. It is thus likely that the absorption features are due to the dynamics related to the cloud dispersal by photoevaporation and radiation stripping.

One common feature in HI 21cm lines at all positions is the asymmetric distribution of HI gas in velocity space seen in the channel maps. Looking at details of HI distribution in velocity, we find abundant neutral hydrogen at LSR velocities larger than -20 km s$^{-1}$ at all 6 positions, while we do not see significant HI emission at the LSR velocity smaller than -20 km s$^{-1}$. We find this behavior at all positions around the CNC. Considering the wide distribution of the HI emission at the same LSR velocity, we think that the HI cloud at $>$-20 km s$^{-1}$ confines the red-shifted portion of the HII region. This agrees with the consistent LSR velocity of the CO components at -18 km s$^{-1}$ across the Tr 14/Carina I region. 

\section{MODELING PDRs AND HII GAS IN THE Tr 14 AND CARINA I REGION} \label{sec:results:pdr}

A number of authors have previously applied PDR models to IR data of the Tr 14 and Carina I region \citep{brooks03, mizutani04, oberst11, okada13, wu18}. Typically, the observations and models included  several of the following: [C\,{\sc ii}] 158 $\mu$m, [O\,{\sc i}] 63, 145 $\mu$m, [C\,{\sc i}] 369, 609 $\mu$m, $^{12}$CO low to mid J transitions, and IR continuum.  Several authors pointed out that the [O\,{\sc i}] 63$\mu$m (and even possibly [C\,{\sc ii}] 158 $\mu$m, see \citealt{mizutani04} and \citealt{wu18}) could suffer self absorption and therefore was not used in the comparisons of observations to their PDR models.  Within a 10 arcminute or roughly 7 pc projected distance from Tr 14, \citet{brooks03} and \citet{oberst11} found rough matches with constant density PDR models that had hydrogen nucleus densities of $\sim$300 -- 3$\times$ 10$^4$ cm$^{-3}$ and FUV fields $G_0$ $\sim$ 600 -- 10$^4$.  {  \citet{kramer08}, using the clumpy KOSMA-$\tau$ model, found $G_0 \sim 500-5\times10^3$ and somewhat higher ensemble average densities of $2\times 10^5$ cm$^{-3}$.} Given that the FUV luminosity of Tr 14 is roughly 2$\times 10^6$ L$_\odot$, this range of $G_0$ corresponds to distances of 2.1 to 8.5 pc if there is insignificant extinction of FUV inside the HII region. The molecular ridge to the southwest of Tr 14 is about 2.3 pc in projected distance, and so the derived $G_0$ values are in rough agreement with the likely geometry of the neutral gas around the Tr 14 HII region.

\citet{wu18} applied constant thermal pressure Meudon PDR models \citep{lepetit06} and found a range of $P_{\rm th}\sim 3\times 10^7$--$3\times 10^8$ K cm$^{-3}$ and $G_0 \sim 3\times 10^3$--$5\times 10^4$, somewhat higher values than previous authors. We discuss these high values below.  Our main thrust in this section is to understand and discuss the interesting relation between the thermal pressure in the PDR with the incident FUV flux found by \citet{wu18}. \citet{wu18} use PACS observations of CO (up to $J$ = 13 -- 12) and both CI fine structure transitions to find best fit PDR models for each pixel in a large map of Car I-E, Car I-S, and Car I/II.  The main free parameters in the models are  $P_{\rm th}$  and $G_0$ (and to a lesser extent the beam filling factor and total column density through the PDR layer) and they use the best fit to each pixel to generate a large number of $P_{\rm th}$ and $G_{0}$ pairs.\footnote{In fact \citet{wu18} used Mathis units and gave their FUV fit in these units, $G_{UV}$. In the Habing units that we use, the relation between the two is $G_0= 1.3G_{UV}$.  We convert in this section the \citet{wu18} results to $G_0$ units.} From the observations and the modeling of each pixel, the empirical relation is
\begin{equation}
P_{\rm th}=2.7\times 10^4 G_{0}^{0.9}\ {\rm K\ cm^{-3}}.
\label{Eq:P_pdr}
\end{equation}
\citet{wu18} do not quote errors in this fit, but their Figure 13 suggests that the errors could be significant.

In this section we first analytically derive the expected relation of the applied pressure to the PDR, $P_{\rm PDR}$,  to $G_0$, using the Str\"omgren relations for HII regions, and the relative strengths of the EUV luminosity $\Phi _{EUV}$  and the FUV luminosity $\Phi_{FUV}$ from the OB association.  We note that, even in steady state, the {\it applied} pressure, $P_{\rm PDR}$, may differ from the {\it thermal} PDR pressure $P_{\rm th}$ because of other sources of PDR pressure support (see below). We compare the relation $P_{\rm PDR}$ to $G_0$ with the \citet{wu18} semi-empirical relation  of $P_{\rm th}$ to $G_0$(equation \ref{Eq:P_pdr}). We apply our PDR models  to the  {  integrated intensities  from} STO2 [C\,{\sc ii}] observations,  as well as the literature values of CO ($J$ = 1 -- 0) \citep{rebolledo16}, CO ($J$ = 4 -- 3 and 7 -- 6) \citep{kramer08}, the CI fine structure lines \citep{kramer08}, [O\,{\sc i}] 63 and 145 $\mu$m \citep{mizutani04,oberst11,wu18}, and the IR continuum \citep{preibisch12} and find $P_{\rm th}$, $G_0$ pairs consistent with observations in a manor similar to \citet{wu18}. We use  [N\,{\sc ii}] observations \citep{oberst11} when available to estimate the [C\,{\sc ii}] emission from the ionized HII gas. We find the [O\,{\sc i}] and [C\,{\sc ii}] lines helpful in constraining our fits, although we place less weight on the [O\,{\sc i}] 63$\mu$m integrated intensity fitting ([O\,{\sc i}] 63 $\mu$m can suffer significant self-absorption) except to ensure that the PDR model intensity is at least as bright as observed. We then also compare our PDR model results with our derived analytical relation.

\subsection{Analytic Derivation of Dependence of $P_{PDR}$ on $G_0$.}

The basic Str\"omgren relation for an HII region with no wind cavity and assuming constant electron density $n_e$inside the HII region is
\begin{equation}
\Phi_{EUV}f_{gas} = {4 \over 3} \pi \alpha_r n_e^2 d^3,
\label{Eq:phi_euv}
\end{equation}
where $\Phi_{EUV}$ is the EUV photon luminosity of the source, $f_{gas}$ is the fraction of the EUV absorbed by recombinations in the HII gas (and not the dust in the HII region), $d$ is the Str\"omgren radius of the HII region (and also the distance from the UV source to a PDR lying just outside the ionization front), and $\alpha_r$ is the recombination coefficient of electrons with protons in the HII region. Using the on the spot assumption, so that only recombinations to the excited levels are counted, we take $\alpha_r= 2.6 \times 10^{-13}$ cm$^3$ s$^{-1}$, assuming the HII region temperature is T= 10$^4$ K. We define a proportional number $f$ such that
\begin{equation}
\Phi_{FUV}=f\Phi_{EUV},
\label{Eq:phi_fuv}
\end{equation}
where $\Phi_{FUV}$ is the FUV photon luminosity in the wavelength range 912-2000 \AA. Typically, for large and young OB associations like Tr 14 with a number of very hot and early type stars, $f\sim 1$. For Tr 14 \citet{smith06} finds $\Phi_{EUV} = 2.2\times 10^{50}$ EUV photons s$^{-1}$ and $L_{FUV}= 2.0\times 10^6$ L$_\odot$. The latter luminosity can be approximately converted to a photon luminosity by assuming the average energy of an FUV photon is 10 eV; $\Phi_{FUV} = 4.8\times 10^{50}$ FUV photons s$^{-1}$.   This makes $f=\Phi_{FUV}/\Phi_{EUV}=2.18$.

The incident FUV flux on the PDR just outside the HII region is then written as
\begin{equation}
G_0= {{\Phi_{FUV} f_{PDR}}\over{4 \pi d^2 F_0}},
\label{Eq:g0}
\end{equation}
where $F_0 \simeq 10^8$ photons cm$^{-2}$ s$^{-1}$ is the FUV flux appropriate for $G_0=1$, and $f_{PDR}$ is the fraction of FUV photons which escape dust absorption in the HII region; $f_{PDR} \simeq f_{gas}$. Here we assume that the PDR is a spherical shell that surrounds the HII region, or is a cloud surface with size greater than the distance $d$ to the UV source.  Once we find $G_0$ and $f_{PDR}$ by comparing PDR models with observations, this equation can be used to determine $d$, the distance of the PDR from Tr 14. The thermal pressure in the HII region is given by
\begin{equation}
P_{HII}=2n_eT,
\label{Eq:p_hii}
\end{equation}
where we have assumed that for each electron, there is one positive charge carrier, either H$^+$ or He$^+$.\footnote{Note that if He is doubly ionized, the factor 2 is slightly smaller, but we shall ignore that small correction.}   Assuming that $T= 10^4$ K in the HII region, we use equations (\ref{Eq:phi_euv}) -- (\ref{Eq:p_hii})  to obtain
\begin{equation}
P_{HII}=2.3\times 10^4 f^{-{3/4}} \Phi_{51}^{-1/4} G_0^{3/4} ~~ {\rm K\ cm^{-3}},
\label{Eq:p_hii_full}
\end{equation}
where $\Phi_{51} = \Phi_{EUV}/ 10^{51}$ photons s$^{-1}$. Note that the dependence of $P_{HII}$ on $G_{0}$ ($P_{HII}\propto G_{0}^{0.75}$) is close to, but not quite the same as the empirical relation ($P_{\rm th}\propto G_{0}^{0.9}$)  given by \citet{wu18}. We also note that the simplest assumption, for a confined HII region with a PDR which surrounds it, is that $P_{HII}=P_{PDR}$. However, the HII region around Tr 14 does not appear confined, but is rather a blister HII region that is expanding away from the GMC. In this case, there is an additional pressure on the PDR caused by the ram pressure of the photoevaporating HII gas off the PDR surface. This additional pressure is of order of the thermal pressure (equation A4 in \citealt{gorti02}), so that  the applied pressure on the PDR $P_{PDR} \simeq 2P_{HII}$, where
$P_{HII}$ is the thermal pressure in the outflowing ionized gas from the PDR surface.

The normalization constant in the $P_{PDR}$ versus $G_{0}$ relation can be compared to the normalization constant in Wu et al if we specify the Smith (2006) values of $f$ and $\Phi_{EUV}$ in the Tr 14  cluster of OB stars, and assume $P_{PDR}= 2P_{HII}$,
\begin{equation}
P_{PDR}= 3.7\times 10^4 G_{0}^{0.75}~~ {\rm K\ cm^{-3}}.
\label{eq:ppdrg0}
\end{equation}
This equation appears very similar to the the empirical relation $P_{\rm th}$ vs $G_0$  found by \citet{wu18}. We need, however, to test it in the region of applicability of the Wu et al. relation. The relation found by \citet{wu18} was for PDRs where $G_{0} \sim 3\times 10^3-5\times 10^4$. The Wu et al. relation (equation (\ref{Eq:P_pdr})) gives $P_{\rm th}= 3.6\times 10^7$ and $4.6\times 10^8$ K cm$^{-3}$ for $G_{0} = 3\times 10^3$ and $5\times 10^4$ respectively, while our analytic relation (Eq.\ref{eq:ppdrg0}) gives $P_{\rm PDR}=1.5\times 10^7$ and $1.2\times 10^8$ K cm$^{-3}$ respectively. Thus, the analytic solution is indeed quite close, perhaps a factor of 3 to 4 lower than the $P_{\rm th}$ found by best fit PDR models in \citep{wu18}.\footnote{We have checked to see if stellar winds could apply sufficient additional pressure to explain the discrepancy, but found, using stellar wind parameters from \citep{smith06}, that the winds are too weak to explain the factor of $\sim 3-4$.}

We again stress here that $P_{\rm th}$ is the thermal pressure in the PDR gas, since PDR models often hold the thermal pressure constant and it is this pressure that PDR modelers, including this paper, plot in their figures. With the exception of the applied pressure $P_{PDR}$ in clumps discussed below, equation \ref{eq:ppdrg0} provides an {\it upper limit} to $P_{\rm th}$ because it assumes that the applied pressure to the neutral region is balanced by the {\it thermal pressure} of the PDR gas. If magnetic pressure supports the PDR gas, then, for a given $G_0$,  $P_{\rm th} < P_{PDR}$, since the sum of thermal pressure and magnetic pressure should equal the applied pressure in steady state. Since the neutral gas around HII regions and in clumps inside HII regions has been pressurized and compressed by the expanding, high temperature HII gas, one might expect larger ratios of  magnetic pressures to thermal pressure in the compressed gas than in ambient gas because magnetic pressure generally increases more rapidly with compression than thermal pressure.

We show below that PDR code provides 10 fits that have filling factors essentially unity, suggesting a shell or partial shell just outside the HII region. However, one fit requires a beam filling factor significantly smaller than unity.  Such a small beam filling factor suggests neutral clumps inside the extended HII region. Opaque clumps are certainly seen in the optical images, clumps appear in the maps of CO we discussed above, and are inferred in the \citet{wu18} study. EUV evaporating clumps inside an HII region have a different relation of $P_{PDR}$ to $G_0$. In fact, another variable is introduced, the radius  $R$ of the clump. If $R>d$, where $d$ is the distance of the clump from the EUV source Tr 14, then the solution given in Eq.\ref{eq:ppdrg0} still applies. However, if $R<<d$, then we have a small EUV evaporating clump inside the extended HII region. A very similar computation to that described above applies, except that the incident EUV flux is absorbed by H atoms that have recombined {\it in the evaporating flow} off of the clump \citep[see ][ for a detailed analysis]{bertoldi90}. Assuming that the flow is ejected at $\sim 10$ km s$^{-1}$, the thermal speed of the $10^4$ K ionized gas at the surface of the clump, one finds, analogous to the steps above for standard HII regions, the applied pressure to the clump is:
\begin{equation}
 P_{PDR} \simeq 4T_{HII}\left({3{F_0 G_0}\over {\alpha_r R f}}\right)^{1/2},
\label{eq:ppdrg1}
\end{equation}
where $T_{ HII} \sim 10^4$ K is the temperature of the ionized gas streaming off the clump, {\it cgs} units are used, and the pressure is in units of K cm$^{-3}$. Note that for a fixed $G_0$, the pressure now depends on the clump radius R.   For $R<<d$ the derived pressure is higher than $P_{PDR}$ given in Eq.\ref{eq:ppdrg0}. Small clumps require higher densities at the ionization front in order to absorb the incident EUV since the characteristic distance R for the EUV be absorbed is smaller. The higher densities produce higher applied pressures to the PDR. It is possible that some of the higher pressures found by \citet{wu18} are caused by clumps along their lines of sight. Alternatively, the area mapped by \citet{wu18} may have localized sources of UV which can lead to higher $P_{\rm PDR}$ for a given $G_0$ (see equation \ref{Eq:p_hii_full}). Another possibility is differences in the chemistry and heating processes in the two PDR codes.

\subsection{PDR Modeling and STO2 [C\,{\sc ii}] Results}

\begin{figure}[h!]
\includegraphics[scale=0.19]{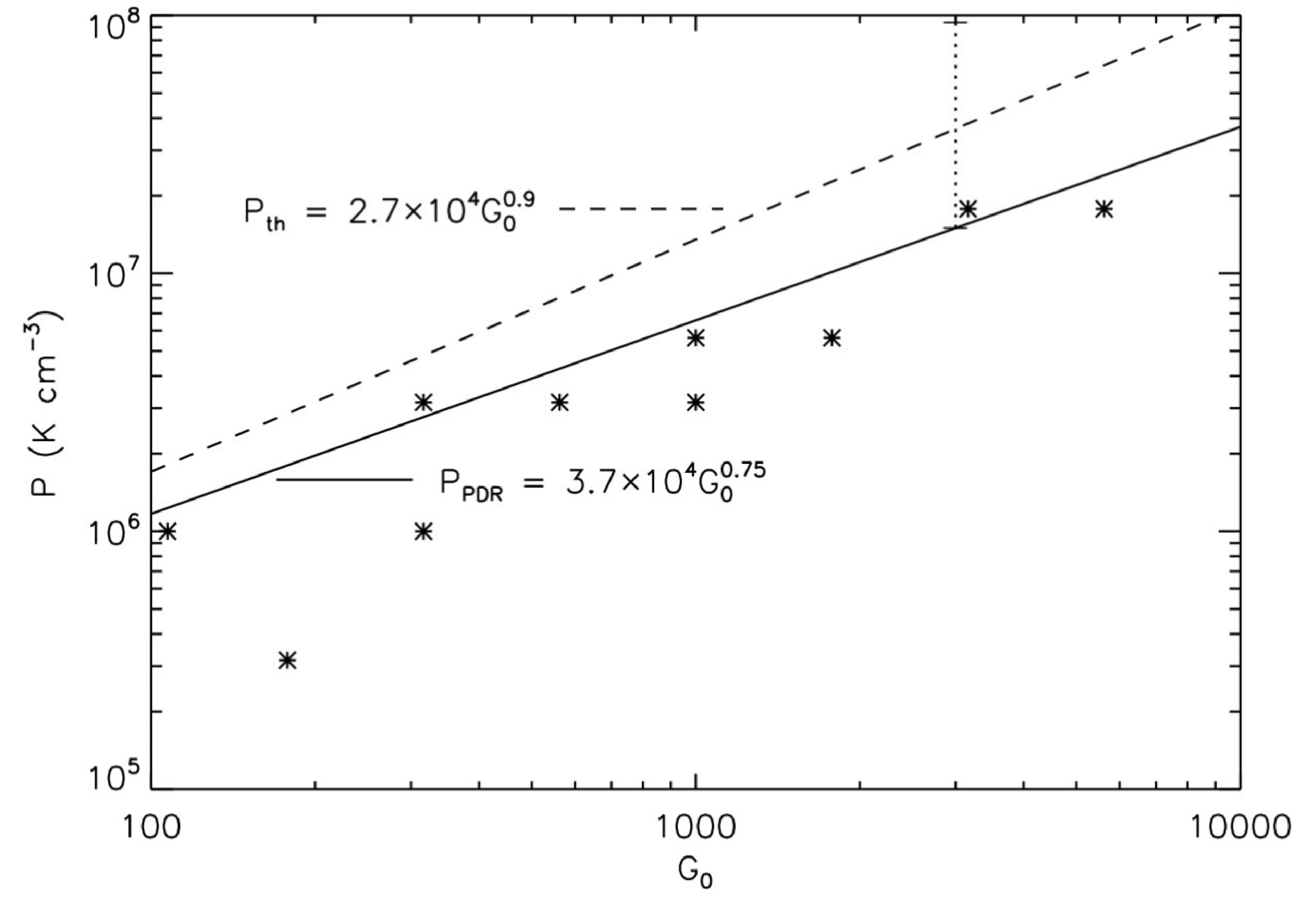}
\caption{The model thermal pressure in the PDR layer, $P_{\rm th}$, versus model incident radiation field $G_0$ is indicated by the star symbols for the 10 features modeled. The alphabets denote the seven line of sights. The solid line is our analytic relation which relates the applied pressure to the PDR, $P_{\rm PDR}$, to $G_0$ for a PDR shell lying just outside the HII region (Eq.\ \ref{eq:ppdrg0}). Results that fall below this line might suggest magnetic pressures that support the PDR, rather than thermal pressure. The dashed line is the relation of PDR thermal pressure to $G_0$ from \citet{wu18}. The dotted line shows how much the predicted applied pressure could be raised if the PDR were a 0.1 pc radius clump inside the HII region.}
\label{fig:pvsg0}
\end{figure}

The analysis of the observations are carried out using a photodissociation region model based on that of \cite{wolfire10} and \cite{hollenbach12} with additional updates noted in \cite{neufeld16}. The models calculate the steady-state chemical abundances and thermal balance gas temperature of a layer of gas of constant thermal pressure, $P_{\rm th}$, exposed to a far-ultraviolet radiation field, $G_0$, and cosmic-ray ionization rate $\zeta_{\rm CR}$. The radiation field is measured in units of the interstellar field of Habing integrated between 6 eV and 13.6 eV ($=1.6 \times 10^{-3}$ ergs ${\rm cm^{-2}}$ ${\rm s^{-1}}$). To aid in the analysis we have carried out a grid of models with $G_0$ varying in log steps of 0.25 between $-0.5 \le \log G_0 \le 6.5 $ and thermal pressure varying in log steps of 0.25 between $2 \le \log P_{\rm th}/({\rm K\, cm^{-3}}) \le 8$.

The cosmic-ray ionization rate and gas-phase abundances are held fixed as given in Table \ref{tab:parameter}. The model output consists of { integrated emission} line intensities  as functions of $G_0$ and $P_{\rm th}$ for lines which arise in the  atomic and molecular gas in the PDR including, [C\,{\sc ii}] 158 $\mu$m, CO  1--0, CO 4--3, CO 7--6, [C\,{\sc i}] 610 $\mu$m, [C\,{\sc i}] 370 $\mu$m, [O\,{\sc i}] 63 $\mu$m, and [O\,{\sc i}] 145 $\mu$m. The PDR models do include self absorption inside the PDR region, as the lines emerge from the illuminated face. However, they neglect self absorption by cool foreground clouds or for lines of sight that approach the PDR from the non-illuminated (back) side.  The [O\,{\sc i}] 63 $\mu$m line has the highest optical depth and is the most prone to this effect. Therefore, in our best fits, we allow the PDR model to somewhat overpredict the {  integrated} [O\,{\sc i}] 63 $\mu$m line. However, in the models presented below, it is never more than a factor of $\sim$2 -- 3.  We emphasize that we do not allow our best-fit models to {\it underpredict} the [O\,{\sc i}] lines by more than a factor of 2. We find that when there is overlap, the PACS observational [O\,{\sc i}] line intensities \citep{wu18}, even when matched to the ISO beam sizes, give higher intensities than those quoted in \citet{oberst11}. Our models suggest the higher values, and we therefore match our models to the {  integrated and convolved} PACS data when it is available.

{  We use the Mopra $^{13}$CO $J$ = 1--0 observations to estimate the total column of the PDR slab for each of our models.   We take the PDR model temperature at $A_V=5$ into the slab to estimate the average temperature of the $^{13}$CO in order to translate the integrated optically thin line emission to a column or a $\Delta A_V(CO)$ of gas that is both molecular H$_2$ and CO. We assume that the isotopic $^{12}$C/$^{13}$C ratio is 60 \citep{szucs14}. The PDR model allows us to compute the $\Delta A_V(atomic)$, which we define as the $\Delta A_V$ from the PDR surface to where the gas is half molecular H$_2$ and half atomic.    Similarly, we obtain $\Delta A_V(dark)$, which we define as the  $\Delta A_V$ from where the gas is half molecular to where the optical depth in the $^{12}$CO line is unity.  The sum of these $\Delta A_V$ is the total $A_V$ of the PDR slab. }

An additional model parameter is the source area filling factor in the telescope beam, $f_{\rm B}$. In practice, we compute $f_{\rm B}$ by comparing our model CO 1--0 intensity to the observed intensity, and then fit all other lines by matching the ratio of the line to CO 1--0.

Another important model constraint comes from the integrated infrared continuum  observations. We use the {\em Herschel} PACS (70 $\mu$m and 160 $\mu$m) and SPIRE (250 $\mu$m, 350 $\mu$m and 500 $\mu$m) observations \citep{preibisch12} convolved to the STO 2 beam size. We fit each SED with a dust optical depth $\tau_{\rm D}$, dust temperature, $T_{\rm D}$, and emissivity index $\beta$ and integrate under the resultant fit to find the integrated continuum intensity, $I_{\rm IR}$. We assume that this IR continuum arises from the sum of contributions from the PDR and the HII region; $I_{\rm IR} = I_{\rm PDR} + I_{\rm HII}$. Using theoretical spectra from O stars \citep{parravano03,malkov07}, we find that approximately half the dust heating in the PDR is from the FUV photons, the rest from photons outside this band. Therefore, $I_{\rm PDR} = 2\times1.6\times 10^{-3} f_{\rm B}G_0/(4\pi)$ erg cm$^{-2}$ s$^{-1}$ sr$^{-1}$. Our best fit models provide $G_0$ and $f_{\rm B}$, and the observations provide $I_{\rm IR}$. We can then estimate $I_{\rm HII} = I_{\rm IR} - I_{\rm PDR}$.

In the model fits we vary only $P_{\rm th}$, $G_0$, and $f_{\rm B}$ although additional factors such as geometry, abundance variations, and depth of the PDR layer may effect the emitted line intensities. In light of these considerations and possible observational errors we assume that fits to the observed line and continuum intensities to within factors of two are considered to be good fits, although we generally find agreement to better than a factor two. {  Below we discuss each fit and the maximum difference between the good fit model and the observations.}

To estimate the [C\,{\sc ii}] line emission from ionized gas we rely on the observed [N\,{\sc ii}] line intensities. The [C\,{\sc ii}] 158 $\mu$m/[N\,{\sc ii}] 205 $\mu$m ratio is weakly dependent on electron density varying between 5.1 at the low density limit and 6.0 at the high density limit, with a minimum of 3.7 at $n_e \approx 50$ cm$^{-3}$. The [C\,{\sc ii}] 158 $\mu$m/[N\,{\sc ii}] 122 $\mu$m ratio has a stronger dependence on $n_e$ varying monotonically between 0.62 (at the high density limit, roughly $n_e> 1000$ cm$^{-3}$) and 9.7 (at the low density limit, roughly $n_e< 1$ cm$^{-3}$). { These ratios are calculated assuming gas phase abundances of C$^+$ and N$^+$ to be 1.6 $\times$ 10$^{-4}$ \citep{sofia04} and 7.5 $\times$ 10$^{-5}$ \citep{meyer97} respectively. The electron collision strengths are from \citet{tayal11} for [N\,{\sc ii}] and from \citet{tayal08} for [C\,{\sc ii}] for T=8000 K gas.} We can estimate the electron density from the [N\,{\sc ii}] 122 $\mu$m/[N\,{\sc ii}] 205 $\mu$m ratio and then obtain the [C\,{\sc ii}] line intensity from either of the [N\,{\sc ii}] lines. This ratio varies from 0.53 in the low density limit ($n_e\la 1$ cm$^{-3}$) to 9.7 in the high density limit ($n_e\ga 1000$ cm$^{-3}$). If either [N\,{\sc ii}] line is not observed, an alternative approach to obtain the electron density is to use the best fit thermal pressure in the PDR. From equation \ref{Eq:p_hii} we find the electron density from
\begin{equation}
n_e = \frac{P_{HII}}{(\rm K\, cm^{-3})}\frac{1}{2T} = \frac{P_{\rm PDR}}{(2\times 2 \times 10,000)}
\label{Eq:ne}
\end{equation}
where we assume a gas temperature in the ionized gas of 10,000 K and $P_{PDR}$ = 2$P_{HII}$.

We have selected 10 spectral features from 7 positions to analyze in detail (see Figures \ref{fig_lines} and \ref{fig_pdr_lines}). These positions are labeled by letters $a-g$ in Figure \ref{fig_lines}. The points generally lie along a line from the north of the Tr 14 cluster to the southwest into the molecular ridge. For each spectrum we integrate the line profiles over one or two likely velocity components and we associate the (spectrally unresolved)  [N\,{\sc ii}] line intensity (i.e. the ionized gas) with the brighter [C\,{\sc ii}] component. Often the {  observed} [C\,{\sc ii}] line intensity in the bright component is brighter than what can be produced by the neutral PDR emission alone, and the expected contribution from the ionized gas is required to match the observations. {  When available, we use the theoretically expected ratio of [N\,{\sc ii}] to [C\,{\sc ii}] to estimate the [C\,{\sc ii}] flux from the HII region.}  We do not apply a filling factor correction to the ionized gas emission, i.e., we assume the ionized gas fills the beam. {  For observations with higher angular resolution than the [C\,{\sc ii}] resolution, we convolve the observations to provide the integrated intensity in the [C\,{\sc ii}] beam.} If two velocity components are present along the line of sight then the observed continuum emission arises from the sum of both components. Using equation (\ref{Eq:g0}), we also estimate a crude distance of the PDR layer, $d$, away from Tr 14 using the model $G_0$, the FUV luminosity estimate of \cite{smith06}, and the IR continuum intensities obtained by model and observation. Here, we assume that $f_{\rm PDR} \simeq I _{\rm PDR}/I_{\rm IR}$. The following gives a summary of our fits for each point. We follow the summary of each position with a short discussion of these results.

a) {\it PDR North of Tr 14 (l,b) = 287.403, -0.537.} The spectra indicate two basic velocity features, one from -11 km s$^{-1}$ to -4 km s$^{-1}$, and one from  -25 km s$^{-1}$ to  -11 km s$^{-1}$. However, the  CO 4--3 and 7--6, and the [C\,{\sc i}] observations suffer baseline problems and are not reliable, except possibly the CO 4--3 transition in the first velocity feature, which is fairly strong. We model this first feature because we have reliable CO 1--0, [C\,{\sc ii}], [O\,{\sc i}], and IR continuum observations. We find an excellent fit (the model fits the observation to within a factor of 1.2 for each line and the continuum) to all of these observations with a PDR model with log $P_{\rm th}= 7.25$, log $G_0 = 3.75$, and  $f_{\rm B} = 0.1$.  {  The $\Delta A_V$ for the atomic, dark and molecular CO gas are 1.1, 2.3, and 4.4 respectively.}
The beam filling factor suggests a clump or clumps in the beam.  If a single clump dominates, then its diameter is about 0.3 of the beam diameter, or radius $R=0.1$ pc.  On the other hand, we may be observing just a portion of a larger cloud. The incident $G_0$ suggests a distance from Tr 14 of about 2.5 pc. This red shifted component compared to the HII gas (whose emission centers at approximately  -17 km s$^{-1}$) may be on the far side of the HII region, traveling away from us.  As we shall demonstrate below, out of the 10 features modeled this is the only spectral feature which suggests a small clump in the beam. As seen in the H$\alpha$ map, this position lies in a direction of unobscured optical emission, and may be primarily ionized gas. This is borne out by the strong [C\,{\sc ii}] emission from the second velocity feature, which likely arises in the ionized gas.

b) {\it The middle point between  Tr 14 and the dust lane (l,b) = 287.392,-0.596}. The spectra suggest a single velocity component -24 km s$^{-1}$ to  -7 km s$^{-1}$. [C\,{\sc i}] observations suffered baseline problems and were not used in this fit, which matched the CO 1--0, 4--3, 7--6 and the two [O\,{\sc i}] lines. The PDR model fit is $\log P_{\rm th} = 5.5$, $\log G_0 = 2.25$, and $f_{\rm B} \simeq 0.5$. {  The $\Delta A_V$ for the atomic, dark and molecular CO gas are 0.6, 2.2, and 1.7 respectively. The ratio of the PDR model flux to the observed flux (henceforth ``m/o") ranges from 0.5 for [O\,{\sc i}] 63 $\mu$m to 1.9 for CO 7--6.} The observed [N\,{\sc ii}] scales to predict a [C\,{\sc ii}] { integrated intensity from the HII region that is a factor of 0.54} of the observed; our PDR model fit only adds about 10\% to the HII contribution, so that the HII region dominates the [C\,{\sc ii}] production in this case. The same is true of the IR continuum emission. We find only 10\% comes from the PDR, and the rest from the HII region. In general, we find that when the HII region dominates the [C\,{\sc ii}], it also dominates the IR, suggesting significant extinction of the UV as it traverses the HII region to the PDR. Taking into account this extinction, we find $d_{\rm PDR} = 4.9$ pc.

c) {\it CG South of Tr 14 (l,b) = 287.386, -0.604.} We fit a single velocity component extending from -24 km s$^{-1}$ to  -7 km s$^{-1}$.  The PDR model fit is $\log P_{\rm th} = 6.0$, $\log G_0 = 2.5$, $f_{\rm B} \simeq 0.5$. {  The $\Delta A_V$ for the atomic, dark and molecular CO gas are 0.5, 2.2,  and 2.4 respectively. The PDR fit is based on the CO, [C\,{\sc i}] and [O\,{\sc i}] fluxes.  For [O\,{\sc i}] 63 and 145$\mu$m  the m/o=1.1 and 1.3 respectively. The worst m/o = 1.8 is for CO 7--6.} Our model fits the [C\,{\sc ii}] if we scale the observed [N\,{\sc ii}] 205 $\mu$m to estimate [C\,{\sc ii}] from the HII region. We find about 90\% of the [C\,{\sc ii}] arises from the HII region, and 10\% from the PDR. We note here that we have found another test besides [N\,{\sc ii}] to determine if the [C\,{\sc ii}] arises mostly from the ionized gas. If the {  flux} ratio [C\,{\sc ii}]/CO(1--0) $\ga 10^4$, then the HII region dominates. PDR dominates typically when the ratio is of order 1000 -- 4000 \citep[see also ][]{wolfire89}. Again, just like the [C\,{\sc ii}], we find that  15\% of the IR comes from the PDR, the rest from the HII region. Using the implied extinction in the HII region, we find $d_{\rm PDR} = 4.5$ pc.

d)  {\it North surface of Carina I-E (l,b) = 287.376,-0.622.} Here the spectra suggest two velocity components. The first component extends from -20 km s$^{-1}$ to  -8 km s$^{-1}$. The PDR model fit is  $\log P_{\rm th} = 6.0$, $\log G_0 = 2.0$, $f_{\rm B} \simeq 0.6$. {  The $\Delta A_V$ for the atomic, dark and molecular CO gas are 0.08, 1.9 and 1.8 respectively. The PDR model fit is based on the CO and [C\,{\sc i}] observations. The worst m/o = 2.0 is for [C\,{\sc i}] 2--1.} The ionized gas is associated with this component and scaling from the [N\,{\sc ii}] we find a match to the [C\,{\sc ii}] from this component. Only about 8\% comes from this PDR component. Similarly, this component only contributes about 3\% of the IR continuum. As noted below, the IR comes from  the other PDR velocity component, with a 10\% contribution from the HII region. Ignoring extinction in the HII region in this case, we find $d_{\rm PDR} = 20$ pc. This gas is somewhat red shifted relative to the ionized gas, and may lie behind the HII region, which apparently extends quite far back in this direction.

The second velocity component extends from -35 km s$^{-1}$ to  -20 km s$^{-1}$. The PDR model fit is $ \log P_{\rm th} = 7.25$, $\log G_0 = 3.5$, $f_{\rm B} = 0.6$, and $d_{\rm PDR} = 3.5$ pc. {   The $\Delta A_V$ for the atomic, dark and molecular CO gas are 0.8, 2.2, and 4.7 respectively. The PDR model fit is based primarily on the CO, [C\,{\sc i}] and [C\,{\sc ii}] fluxes, but the [O\,{\sc i}] fluxes} help drive the fit to high $P_{\rm th}$  and $G_0$. {  The m/o flux ratios vary from 0.63 for [C\,{\sc i}] 2--1 to 1.4 for CO 7--6, while the more suspect [O\,{\sc i}] lines have m/o = 3.7 and 0.64 for 63 and 145$\mu$m respectively.}
Since the 63 $\mu$m line can be self absorbed, as discussed, we discount the mismatch to the 63 $\mu$m line. We cannot get higher values for the 145 $\mu$m line without ruining the fits to the CO and [C\,{\sc i}] lines. Essentially all the [C\,{\sc ii}] and IR emission arise from the PDR. This blue shifted gas is likely on the near side of the HII region, moving toward us. {  We note that this velocity component dominates the line intensity integrated over all velocity and that this position is close to that modeled by Kramer et al (2008). Our model result for $G_0$ is in exact agreement with Kramer et al., but our density at $A_V=1$ is $6\times 10^4$ cm$^{-3}$ , whereas they obtained $2\times 10^5$ cm$^{-3}$ in their constant density clumpy model.}

e) {\it The Oberst et al. Car I position (l,b) = 287.370,-0.630.} The spectra suggest two velocity components. The first extends from -20 km s$^{-1}$ to -7 km s$^{-1}$. The PDR model fit is $\log P_{\rm th} = 6.5$, $\log G_0 = 2.5$, and $f_{\rm B} \simeq 0.5$. {  The $\Delta A_V$ for the atomic, dark and molecular CO gas are 0.2, 2.1, and 2.4 respectively. The worst m/o=2.0 for the [C\,{\sc i}] 2--1 line. The other 4 lines have m/o from 0.7 to 1.5.} The ionized gas is associated with this component, and the [C\,{\sc ii}] in this velocity range must come from the ionized gas, as the PDR component only contributes roughly 5\%. The scaled [N\,{\sc ii}] lines support this, if we use the electron density estimated from the thermal pressure. Likewise, the [C\,{\sc ii}]/CO 1--0 {  flux} ratio is high, $1.7\times 10^4$, suggesting [C\,{\sc ii}] from the HII region. Similarly, this PDR component only supplies about 6\% of the IR continuum. Taking extinction in the HII region into account, the distance to PDR is $d_{\rm PDR} = 3.4$ pc.

The second velocity component extends from -35 km s$^{-1}$ to -20 km s$^{-1}$. The PDR model fit is $\log P_{\rm th} = 6.75$, $\log G_0 = 3.25$, and $f_{\rm B} \simeq 0.8$, {  based on CO, [C\,{\sc i}], [C\,{\sc ii}], [O\,{\sc i}], and IR continuum observations. The $\Delta A_V$ for the atomic, dark and molecular CO gas are 0.8, 2.2, and 4.3 respectively. The fits to the CO and [C\,{\sc i}] lines have m/o that range from 1.0 for CO 1--0 and 4--3 and [C\,{\sc i}] 2--1 to 1.4 for CO 7--6.} As in case d above, our fit to the [O\,{\sc i}] lines match the 145 $\mu$m line satisfactorily (m/o= 0.7), but the model overestimates the 63 $\mu$m line by a factor 2.5. The latter misfit may be  caused by self absorption.\footnote{This feature and the second velocity feature in d are the only ones where we could not match the [O\,{\sc i}] lines to better than a factor of two.} The PDR model fit provides $\sim 90$\% of the [C\,{\sc ii}] emission, so little arises from the ionized gas in this case. This PDR velocity component supplies roughly 0.5 of the observed IR emission. Taking into account moderate extinction in the HII region, we find $d_{\rm PDR} \simeq  3.4$ pc. This blue shifted gas likely lies in the foreground, between the observer and the HII region.

f) {\it South shell of Carina I-E (l,b) = 287.370,-0.636}. The first velocity component extends from -20 km s$^{-1}$ to  -8 km s$^{-1}$. The PDR model fit is $\log P_{\rm th} = 6.75$, $\log G_0 = 3.0$, and $f_{\rm B} \simeq 0.5$, {  based primarily on fits to the CO and [C\,{\sc i}] lines but also considering  the [O\,{\sc i}] lines (see below).  The $\Delta A_V$ for the atomic, dark and molecular CO gas are 0.5, 2.2, and 3.1 respectively. The m/o ranges from 1.0 for CO 1--0 and 4--3 to 1.8 for [C\,{\sc i}] 2--1.} The observed [N\,{\sc ii}] 205 $\mu$m scales to predict [C\,{\sc ii}] from the HII region that is 70\% of the observed. The PDR model only supplies about 10\% of the observed emission. Therefore, the HII region dominates the [C\,{\sc ii}] production. This PDR component supplies only 17\% of the observed IR continuum.  We show below that the other PDR velocity component contributes about 31\% and thus the HII region contributes about 52\% of the IR continuum. We find then that the first velocity component lies roughly $d_{\rm PDR} = 4.6$ pc from Tr 14. The PDR gas may be somewhat red-shifted with respect to the HII gas, and therefore may lie somewhat behind UV source Tr 14.

The second velocity component extends from -35 km s$^{-1}$ to  -20 km s$^{-1}$. The PDR model fit is $\log P_{\rm th} = 6.5$, $\log G_0 = 3.0$, and $f_{\rm B} \simeq 0.9$, {  based on CO, [C\,{\sc i}], [C\,{\sc ii}] and to some extent [O\,{\sc i}] and IR continuum. The $\Delta A_V$ for the atomic, dark and molecular CO gas are 0.7, 2.2, and 3.4 respectively.  The m/o for CO and [C\,{\sc i}] range from 1.0 for CO 1--0 and 4--3 to 1.4 for CO 7--6. The models predict about equal amounts of [O\,{\sc i}] emission from each velocity component. When summed the m/o = 2.7 for 63$\mu$m and 0.85 for 145$\mu$m.}  Most of the [C\,{\sc ii}] in this component comes from the PDR; PDR model predicts 0.83 of the observed [C\,{\sc ii}]. In addition, about 31\% of the IR continuum comes from this component. The derived distance to Tr 14 is $d_{\rm PDR} = 4.9$ pc. The blue shift suggests this PDR is on the near side of the HII region.

g) {\it West shell of Carina I-E (l,b) = 287.355,-0.639.} We fit a single velocity component from -30 km s$^{-1}$ to -10 km s$^{-1}$. Here, we have no [O\,{\sc i}] observations to guide us, but we get an extremely good fit with $\log P_{\rm th} = 6.5$, $\log G_0 = 2.75$, and $f_{\rm B} \simeq 1.7$. {  The $\Delta A_V$ for the atomic, dark and molecular CO gas are 0.4, 2.1, and 2.9 respectively.}  The three CO lines and the two [C\,{\sc i}] lines are all fit to within a factor of 1.05. However, the PDR model fit only provides 28\% of the observed [C\,{\sc ii}] emission, and 31\% of the IR continuum, suggesting that both [C\,{\sc ii}] and IR arise mostly from the HII region. Using the inferred extinction in the HII region, we then estimate $d_{\rm PDR} \simeq 4.8$ pc. The blue shifted velocity range suggests the PDR lies on the near side of the HII region.

Figure \ref{fig:pvsg0} plots pressure $P$ versus $G_0$. The stars show our PDR code fits of the PDR thermal pressure $P_{\rm th}$ and $G_0$ for 10 spectral features along 7 sight lines. The solid line shows the analytic relation of the applied pressure $P_{\rm PDR}$ to $G_0$, for the case where the size scale of the PDR (or clump radius $R$) is large compared to the distance to the UV source. Best fits that fall below this line suggest that magnetic pressure helps support the PDR, and that the applied pressure is matched by the sum of magnetic and thermal pressure. Seven of the 10 fits fall below but within a factor of $\sim 2$ of the applied pressure line. These suggest magnetic pressure is only comparable to or smaller than the thermal pressure. Two points fall slightly (much less than a factor of two) above the relation.  Within the errors, these are consistent with thermal pressure dominating.  However, one fit (the middle point between Tr 14 and the dust lane) lies a factor of six below the applied pressure line. This extreme case may indicate errors in observation (the observations indicate an extremely low CO 7--6/4--3 ratio of $\sim 0.1$, much lower than the other nine observations). However, taking the observations and model fit at face value, it implies magnetic pressure five times the thermal pressure.

The dotted line at $G_0= 3000$ indicates how much the applied pressure would rise if the PDR were a clump with radius R= 0.1 pc that is much smaller than the distance to the source ($d \sim 5$ pc for this FUV field). There are two motivations for showing this result.  The first is that the dashed line shows the relation of $P_{\rm th}$ to $G_0$ found by the PDR modeling of \citep{wu18}. Their fits fall factors of 2--4 above the analytic relation for a shell or for clumps with size larger than $d$. One explanation is that they more often see small clumps in their observations than we do along our lines of sight. The dotted line shows that clumps of size $\sim 0.1$ pc would  raise the applied PDR pressure by a factor of about 6 for a given $G_0= 3000$. The other motivation is that we have one line of sight (a, north of Tr 14) which is the only one in our analysis that appears to be a clump with size smaller than $d$. As noted above, its beam filling factor suggests a clump of radius $R\simeq 0.1$ pc. If this is the case, note that our fit falls below the predicted applied pressure on this clump. This suggests magnetic support for this clump. Indeed, one can compute that if this clump were solely thermally supported, it would gravitationally collapse in a time much shorter than the $\sim 1$ Myr life of the HII region/PDR complex. Therefore, it {\it must} be dominated by magnetic pressures.

We conclude this subsection by summarizing the main points of the PDR modeling. PDRs are observed at both blue shifts and red shifts from the HII region gas, suggesting both foreground and background PDRs. The PDRs on the molecular ridge may have a more edge-on geometry. The distances of the PDRs from the HII region generally are somewhat greater than 2.3 pc, which is the projected distance of the molecular ridge from the Tr 14 OB association. This seems reasonable given the three dimensional aspect of this blister HII region and the neutral gas that surrounds it.

Of 10 spectral features analyzed along seven lines of sight ({\it los}), we find six spectral features in which the [C\,{\sc ii}] and IR continuum mainly come from the HII region, and four where both arise mostly from the PDR. If unresolved spectra had been used, so that we could not separate features along a los, we find that out of the 7 los, three would have [C\,{\sc ii}] dominated by PDR emission, two by HII emission, and two with comparable contributions. As would be expected, the PDRs dominate when the los are directed at the molecular ridge. It is noteworthy that [C\,{\sc ii}] and IR correlate in this way: if one comes primarily from the HII region, then so does the other. Intuitively, this makes sense. High dust extinction in the HII region leads to much lower FUV fields incident on the PDR, and thus less [C\,{\sc ii}] emission. Like other authors \citep[e.g.][]{oberst11} we find that [C\,{\sc ii}] often has a significant contribution from the ionized gas in this region around Tr 14.  {  Summing the [C\,{\sc ii}] luminosities over the seven {\it los}, we find that 3.7 times more [C\,{\sc ii}] luminosity comes from the HII region than from the PDR.} The edge on geometry of the molecular cloud with respect to Tr 14 and the blister geometry leading to expanding HII gas toward and away from us may help explain the importance of the HII region. The high EUV luminosity of Tr 14 leads to larger columns in the HII gas and therefore stronger [C\,{\sc ii}] emission from the ionized gas, and  also contributes to the possibility of significant dust extinction in the HII region and therefore dominant contributions to the IR continuum.

{  The PDR model fits provide a measure of the mass in atomic gas, dark gas (H$_2$ but little CO), and CO gas.  Summing the 10 PDRs found on these seven sight lines, we find a beam averaged mass proportion going as  1:4.1:5.6  for atomic:dark:CO. If these 10 regions are representative, this gives an estimate of the mass budget in this Tr 14 region.   }

The correlation of the thermal pressure in the PDR with the incident FUV field, observed by \citet{wu18} and confirmed by our PDR models presented here, finds some theoretical basis using a simple analytic model of evaporating ionizing gas pressurizing the PDR surface of a molecular cloud or clump.  The good correlation of observation and theory suggests that our PDR models are reasonably correct.

\begin{deluxetable}{lc}
\tablecolumns{2}
\tablewidth{0pt}
\tablecaption{Model Parameters\label{tab:parameter}}
\tablehead{
\colhead{Parameter$\,\,\,\,\,\,\,\,\,\,\,\,\,\,\,\,\,\,\,\,\,\,\,\,\,\,\,$}
     & \colhead{Value}
}
\startdata
C/n\tablenotemark{a} & $1.6\times 10^{-4}$  \\
O/n\tablenotemark{a} & $3.2\times 10^{-4}$ \\
Mg/n\tablenotemark{a} & $1.1\times 10^{-6}$ \\
Si/n\tablenotemark{a} & $1.7\times 10^{-6}$ \\
Fe/n\tablenotemark{a} & $1.7\times 10^{-7}$ \\
S/n\tablenotemark{a} & $2.8\times 10^{-5}$ \\
$\delta v_{\rm D}$\tablenotemark{b} & 1.5 km ${\rm s^{-1}}$ \\
$\zeta_{\rm CR}$\tablenotemark{c} & $2\times 10^{-16}$ ${\rm s^{-1}}$\\
$A_{\rm V}$\tablenotemark{d} & 10\\
\enddata
\tablenotetext{a}{Gas phase abundance per hydrogen nucleus}
\tablenotetext{b}{Doppler line width}
\tablenotetext{c}{Primary cosmic-ray ionization rate per hydrogen, \citealt{indriolo15,neufeld17}}
\tablenotetext{d}{Depth of PDR layer}
\end{deluxetable}

\section{DISCUSSION} \label{sec:dis}

\subsection{Three Dimensional Morphology of the Tr 14/Carina I Region} \label{sec:dis:3d}

\begin{figure*}
\centering
\includegraphics[angle=0,scale=0.3]{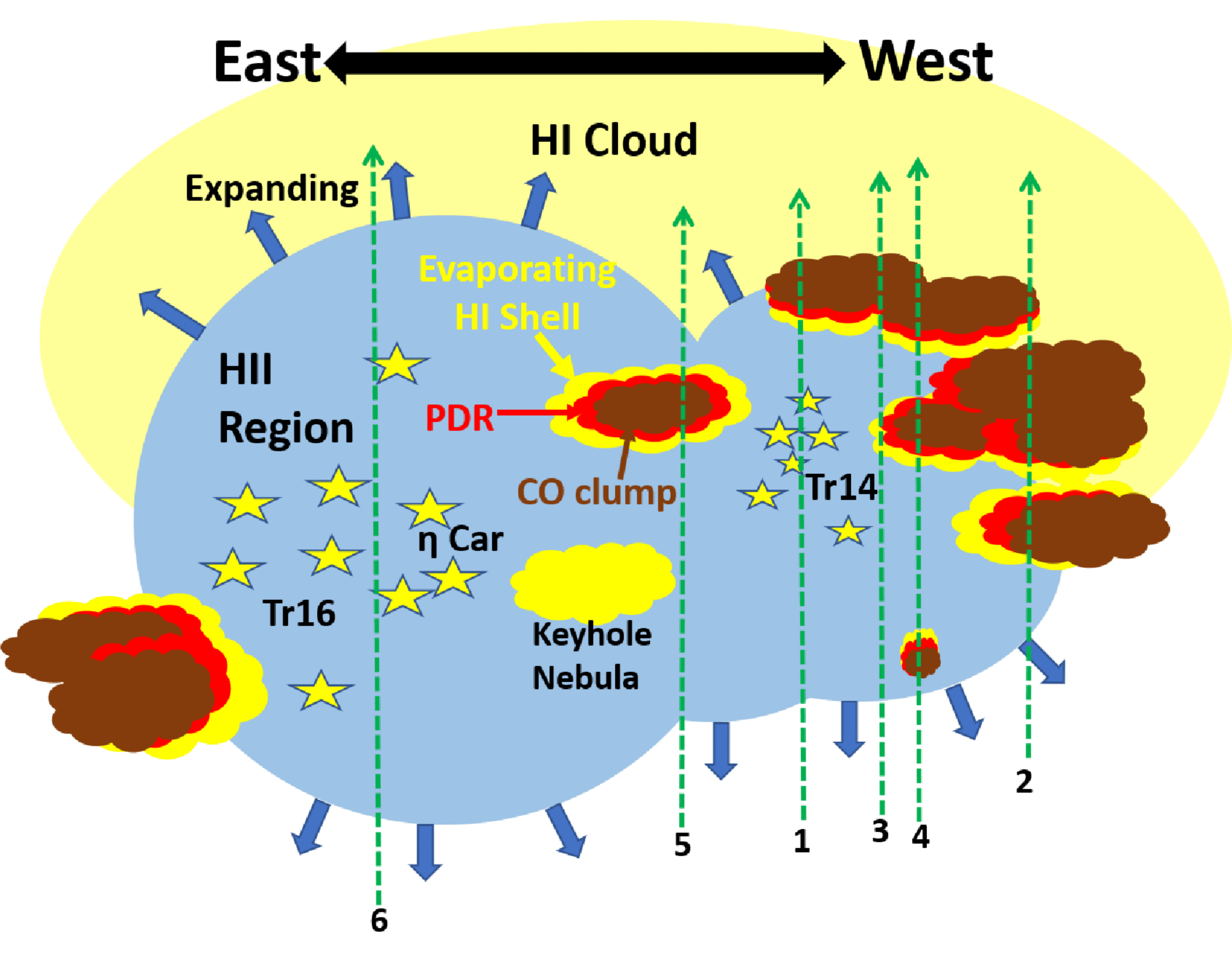}
\caption{Schematic picture of the Carina Nebula Complex. The lines of sight (los) towards the 6 positions are denoted by the green dashed arrows with the position numbers. The green dashed arrows indicate the los directions from the observer.}
\label{schematics}
\end{figure*}

We show a schematic picture of the Tr 14/Carina I region in Figure \ref{schematics} illustrating the following six findings about the three-dimensional morphology of the Tr 14/Carina I Region:

1. The HII region in the Tr 14/Carina I is expanding and is likely to be asymmetric where the red-shifted portion of the HII region is confined by the HI cloud while the blue-shifted portion of the HII region is freely expanding toward us. We find that the H92$\alpha$, H$\alpha$, and [N\,{\sc ii}] 6548$\rm \AA$ lines near Tr 14 and Tr 16 typically have highly asymmetric profiles with long tails to negative velocities relative to the CO clumps around -20 km s$^{-1}$ (e.g., the spectra towards positions 1, 5, and 6 in Figure \ref{fig_lines}). The HI emission in velocity space over the entire CNC is observed mainly at LSR velocities greater than -20 km s$^{-1}$, while we do not find significant HI emission at LSR velocities lower than -20 km s$^{-1}$, indicating that the HI gas impedes the HII region from expanding away from us.

2. The CO and HI clumps partially confine the HII regions of Tr 14 and Tr 16. The channel maps in Figures \ref{fig_chan1}--\ref{fig_chan4} show that molecular and HI clouds/clumps partially surround Tr 14 and Tr 16 in PPV space. We see that there are CO clouds/clumps surrounding the north, the west, and the south of Tr 14, and that there are CO clouds/clumps surround the east of Tr 16.  

\begin{figure*}
\centering
\includegraphics[angle=0,scale=0.31]{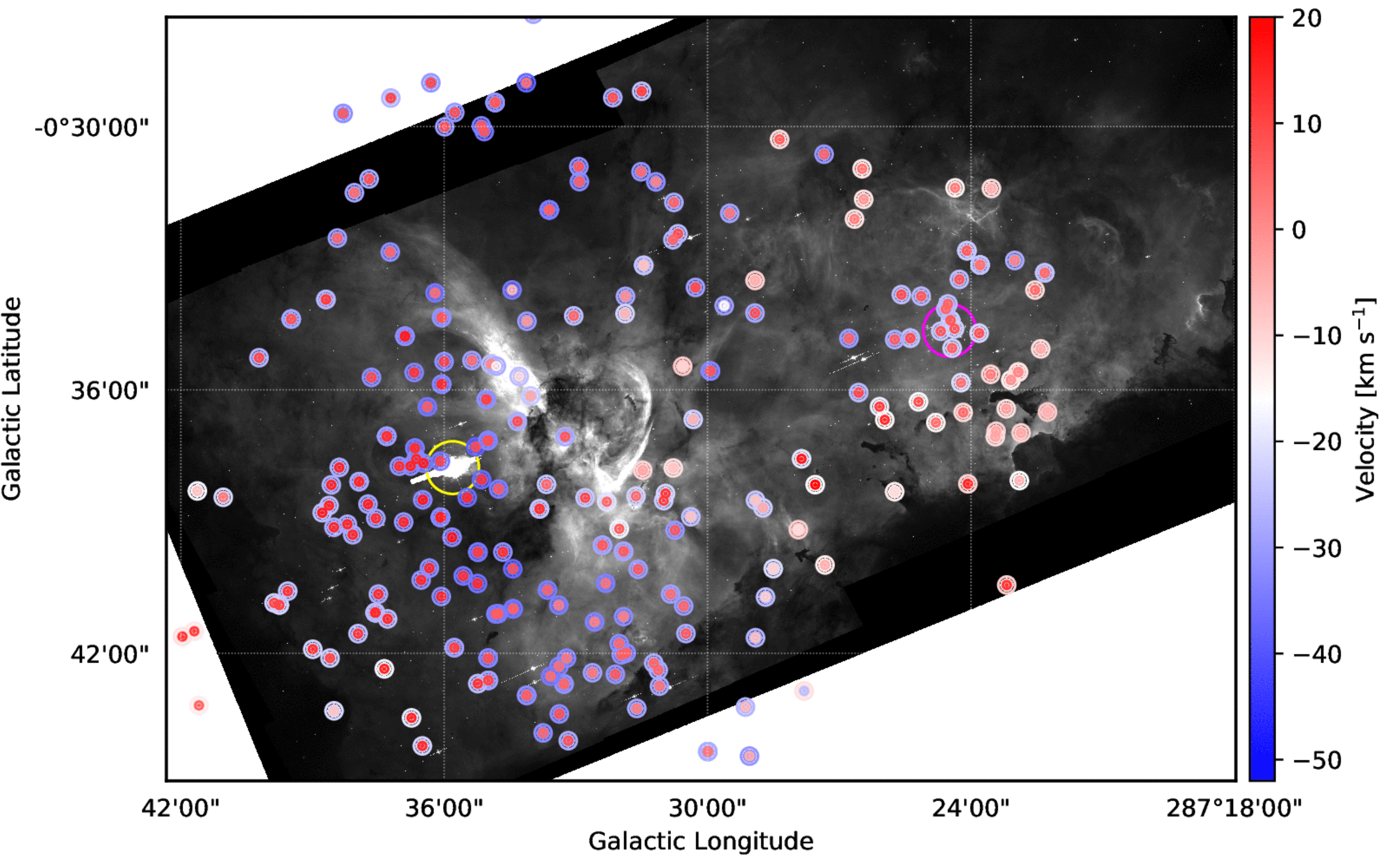}
\caption{LSR Velocity of the double peaks of the H$\alpha$ lines observed using {\it Gaia} \citep{damiani16}. The background image is the H$\alpha$ observed using {\it Hubble} \citep{smith06}. Each inner disk denotes the LSR velocity of the red-shifted intensity peak, and each outer ring denotes the LSR velocity of the blue-shifted intensity peak. The velocity difference between the red- and blue-shifted peaks shows the size of the HII region along the line of sight.}
\label{ionized_gas_vel}
\end{figure*}

3. The Tr 14 and Tr 16 clusters have made their own HII regions that have partially merged as they expanded. Figure \ref{ionized_gas_vel} shows the LSR velocities of the red- and blue-shifted peaks of H$\alpha$ lines observed using {\it Gaia} \citep{damiani16}. The small separation between the red- and blue-shifted peaks indicates that there are either HI or CO clumps that confine the HII regions or the edge of the HII bubble. We find that the separation is reduced between Tr 14 and Tr 16. In the HI 21cm map, we see there are numerous HI clumps between Tr 14 and Tr 16 separating the HII regions of the two clusters.

4. The HII region of Tr 16 is likely more extensive than that of Tr 14. A large separation between the red- and blue-shifted peaks of the H$\alpha$ lines (Figure \ref{ionized_gas_vel}) indicates a faster expansion of the HII region in Tr 16 compared to Tr 14. The expansion speed increases with the radial distance from the ionizing source \citep[e.g.,][]{krumholtz07}, thus a larger HII region is expected to have a faster expansion speed. We found that Tr 16 has roughly three times larger projected area and a factor of two larger velocity separation between two intensity peak compared to Tr 14, suggesting that the HII region of Tr 16 is significantly larger than Tr 14 along the line of sight as well. In addition, CO clouds still confine Tr 14 while Tr 16 seems to have cleared out the nearby dense structures except the east side.

5. CO clouds/clumps have PDRs which show a hint of cloud dispersal by photoevaporation and radiation stripping. We find that most of the [C\,{\sc ii}] emission is spatially correlated with the CO clouds/clumps, but the intensity peaks of the [C\,{\sc ii}] emission are slightly displaced from the CO intensity peaks. In the channel maps, we also found that the morphology of the [C\,{\sc ii}] emission often has arc-shaped clumps surrounding or a strip adjacent to a CO cloud, suggesting there are PDRs and ionization fronts on many of the CO clouds/clumps. Toward dense clumps (e.g., Carina I-E), most of the HI spectra show absorption features (e.g., positions 3 and 4 in Figures \ref{fig_lines}). HI absorption features either red- or blue-shifted relative to $^{12}$CO emission are likely due to photoevaporation and radiation stripping of the exposed CO clumps as predicted from theories \citep[e.g.][]{bertoldi89,bertoldi90,lefloch94,mellema98,mcleod16}. This suggests that dense CO clumps subjected to strong external radiation show a multi-layered structure. 

6. Some of the CO clumps are located within the HII regions. The locations of the ionized, neutral, and molecular gas along the line of sight are not trivial to determine, but we think that the HI absorption features and the [C\,{\sc ii}] emission provide information about the locations of CO clumps relative to the HII regions. Near Tr 14, most of the HI spectra shows both red- and blue-shifted absorption features with respect to CO clumps at $\sim$-20 km s$^{-1}$ (see Figures \ref{fig_chan1}--\ref{fig_chan4}). If massive stars powering  HII regions are in front of and behind the CO clouds/clumps along the line of sight we may observe HI absorption features red- and blue-shifted relative to the CO clouds/clumps originating from their evaporating surfaces. The strongest double absorption features are found near Carina I-E and the Keyhole Nebula suggesting that they are cold dense clumps or pillars located within the HII regions.

The schematic in Figure \ref{schematics} { presents} the six findings on the ISM structure in the Tr 14/Carina I region in addition to descriptions of Tr 16 in \citep{brooks00}. The HII region is mostly open to us, and the CO clumps are within the HII region or partially confine the HII region. The CO clumps exposed to the HII regions have layered structures with evaporating surfaces. There is an HI cloud enclosing  and confining the far-side/red-shifted portion of the HII region while the near/blue-shifted side is expanding more freely, similar to the {\it Champagne} model \citep{tenorio-tagle79} but with some of dense CO clumps being still left within the HII regions.

\subsection{EUV Photoevaporation of the GMC}

The GMC which mostly lies to the west of Tr 14 has an estimated mass of about $\sim 10^6$ M$_\odot$ \citep[e.g.,][]{grabelsky88,smith07}. \citet{grabelsky88} also estimate the radius of the GMC to be $R_{GMC}\sim 66$ pc. \citet{smith06} estimates the total present-day EUV photon luminosity produced by the 65 O stars and the three WNL stars in Carina to be $\Phi_{EUV}\sim 10^{51}$ EUV photons s$^{-1}$ \citep[see ][]{smith07}. Of this, about $6\times 10^{50}$ EUV photons s$^{-1}$ come from Tr16 and $2.2\times 10^{50}$  EUV photons s$^{-1}$ come from Tr 14. Our STO-2 region mostly encompasses the region illuminated by Tr 14, but we examine both photoevaporation from our smaller region and also consider the photoevaporation caused by both Tr 14 and Tr 16 as they illuminate the GMC.

We consider a blister configuration for the GMC; that is, Tr 14 and Tr16 are assumed at some distance from the surface of the GMC but at a distance small compared to the radius of the GMC, $R_{GMC}$, as well as to the radius R of the illuminated surface. The EUV-induced mass loss rate from a surface area of radius $R$ is given
\begin{equation}
\dot M_{evap}\simeq \pi R^2 n_e \bar m v_{evap},
\label{Eq:m_evap}
\end{equation}
where $n_e$ is the typical electron density at radius $R$, $\bar m\sim 2\times 10^{-24}$ gm is the gas mass per electron, and $v_{evap}\sim 10$ km s$^{-1}$ is the speed at which the ionized gas moves off the surface of the GMC.

Using equation (\ref{Eq:phi_euv}), we derive
\begin{equation}
n_e(R) \simeq \left({{3 \Phi_{EUV}}\over{4 \pi \alpha_r R^3}}\right)^{1/2} \simeq 30\Phi_{51}^{1/2}R_{20}^{-3/2} \ {\rm cm^{-3}},
\label{Eq:n_e_evap}
\end{equation}
where $R_{20}\equiv R/10^{20}$ cm and $\Phi_{51}\equiv \Phi_{EUV}/10^{51}$  EUV photons s$^{-1}$. Note that $R$ here is really the distance from the ionizing source to the GMC cloud surface. This is somewhat larger than the radius $R$ of the illuminated surface. However, because we assume the distance of the stars from the cloud surface is small compared to $R$, these two distances are approximately the same.

Substituting equation (\ref{Eq:n_e_evap}) into equation (\ref{Eq:m_evap}), we obtain
\begin{equation}
\dot M_{evap} \simeq 3\times 10^{-2} R_{20}^{1/2} \Phi_{51}^{1/2}\ {\rm M_\odot \ yr^{-1}},
\label{Eq:m_evap2}
\end{equation}
Note that although the GMC surface very close ($R \sim 3$ pc) to Tr 14 and Tr 16 is illuminated with a higher EUV flux and produces a higher mass {\it flux} loss, the more distant ($R\sim 30$ pc) regions of the GMC actually dominate the evaporative mass loss from the cloud, because they have so much more area.

The existence of a large area of the cloud being illuminated and ionized by the massive stars in Tr 14 and Tr 16 has been observed by numerous investigators. \citet{mizutani04} detect fine structure IR lines of [O\,{\sc iii}], [N\,{\sc iii}], and [N\,{\sc ii}] and infer the existence of diffuse ($n_e < 100$ cm$^{-3}$) ionized gas over at least 30 pc. Considering just the region dominated by Tr 14, we see from equation (\ref{Eq:n_e_evap}) that although the electron density at the GMC surface at $R\simeq 3$ pc is $n_e \simeq 450$ cm$^{-3}$, the electron density on the GMC surface at $R\simeq 15$ pc is $n_e \simeq 40$ cm$^{-3}$, in accordance with the results of \citet{mizutani02}.

If we substitute $\Phi_{EUV} \sim 10^{51}$ s$^{-1}$ for the total of Tr 14 and Tr 16, and also estimate that their EUV flux covers $R \sim 10^{20}$ cm $\simeq 30$ pc of the GMC surface, then we obtain $\dot M_{evap} \simeq 3\times 10^{-2}$ M$_\odot$ \ yr$^{-1}$.  If the GMC mass is $M_{GMC} \simeq 10^6$ M$_\odot$, then the lifetime of the cloud is $t_{GMC} \equiv M_{GMC}/\dot M_{evap} \simeq 30$ Myr.   This is close to estimated lifetimes of GMCs.  Note, however, that this assumes that the cloud is illuminated by $\Phi_{EUV} \sim 10^{51}$ s$^{-1}$ for the entire 30 Myr, which implies that other associations will form to replace Tr 14 and Tr 16, once their massive stars die in $\sim 3-5$ Myr.  If we just use the $\Phi_{EUV} \simeq 2.2\times 10^{50}$ s$^{-1}$ of Tr 14 and just consider the region mapped by STO2, which is $R\sim 5$ pc, then the evaporative mass loss rate from this region is only $\dot M_{evap} \simeq 5\times 10^{-3}$ M$_\odot$\ yr$^{-1}$.  We will compare this below with the neutral mass loss  caused by the expansion of the HII region in this region.

\subsection{Mass Loss from the GMC Due to Expulsion of Neutral Clumps or Shells}

The escape speed $v_{esc}$  of a shell or clump of mass $m$ from a GMC is given:
\begin{equation}
{1\over 2}m v_{esc}^2 = {{G m M_{GMC}}\over R_{GMC}}.
\end{equation}
Thus,
\begin{equation}
v_{esc}\simeq 16\left( {M_6 \over R_{20}}\right)^{1/2} \ {\rm km \ s^{-1}},
\end{equation}
where $M_6\equiv M_{GMC}/10^6$ M$_\odot$.   Assuming $M_6 \simeq 1$ and $R_{20} \simeq 2$ \citep{grabelsky88,smith07}, we find $v_{esc} \simeq 11$ km s$^{-1}$. Assuming the line of sight (los) rest velocity of the GMC is -17 km s$^{-1}$ \citep{brooks03}, we then associate all observed CO, [C\,{\sc ii}] and HI that is traveling at los speeds $v < -23$ km s$^{-1}$ or $v> -11$ km s$^{-1}$ as being liberated from the gravitational potential of the GMC.

We proceed to make a very rough estimate of the mass of neutral gas that is not bound to the GMC, and therefore may be considered to be evaporating off the cloud, perhaps pushed by the expanding HII gas.  First we estimate the mass in neutral gas that is not traced by CO but by the [C\,{\sc ii}] from the PDRs. This includes atomic PDR gas, but also gas that is H$_2$ but has little CO due to photodissociation, sometimes called the hidden molecular gas or the dark gas. We assume that most of the [C\,{\sc ii}] emission comes from the PDRs and not from the HII ionized gas, and that the [C\,{\sc ii}] emission is optically thin. This probably gives a lower limit to the mass since the PDR models show the [C\,{\sc ii}] often has a line optical depth of order unity, but only a lower limit by a factor then of a few. We also assume the PDR densities are sufficiently high that the level populations are in LTE and that the temperature of the PDR is $T>92$ K (as the PDR models suggest).  If the densities are lower or the temperature is lower, then again our estimate is a lower limit.  With these caveats, we then obtain a mass of escaping gas (hydrogen and helium included) that is directly related to the [C\,{\sc ii}] luminosity from the whole region mapped by STO2.

\begin{equation}
M= 1.2 \left({L_{[C_{\rm II}]} \over 1\ L_\odot}\right)\ {\rm M_\odot}
\end{equation}
The observed STO2 [C\,{\sc ii}] luminosity from the los velocity range -40 km s$^{-1}$ to -23 km s$^{-1}$ is $L_{[C_{\rm II}]}= 583$ L$_\odot$. This corresponds then to $707$ M$_\odot$ of atomic and H$_2$ gas (but no CO gas). The observed STO2 [C\,{\sc ii}] luminosity from the los velocity range -11 km s$^{-1}$ to 0 km s$^{-1}$ is $L_{[C_{\rm II}]}= 205$ L$_\odot$. This corresponds then to $246$ M$_\odot$ of atomic and H$_2$ (but no CO) gas. So in sum we find a lower limit of about 1000 M$_\odot$ of evaporating or non-bound gas traced by [C\,{\sc ii}].

{ We use the $^{13}$CO $J$ = 1--0 luminosity to estimate the mass of the evaporating gas traced by CO. This gas is almost certainly optically thin in this isotopic line, and so the mass derives directly
from the luminosity in the line, the temperature of the emitting gas, and the assumption that the gas density is above the critical density, which the PDR models confirm.   }
\begin{equation}
M= 9.6 \times 10^4 \left({T \over 30\ {\rm K}}\right)\left({L_{^{13}CO} \over 1\ {\rm L_\odot}}\right)\ {\rm M_\odot}
\end{equation}
{ In the 10 spectral features that we modeled in detail, the average temperature of the gas at $A_V=5$, typical of the $^{13}$CO  gas, was 22 K, with a range of 18 K to 33 K. The observed Mopra $^{13}$CO luminosity from the los velocity range -40 km s$^{-1}$ to -23 km s$^{-1}$ is $L_{^{13}CO}= 0.0077$ L$_\odot$. These numbers then correspond  to  roughly $540$ M$_\odot$ of H$_2$ gas associated with this blue CO component. The observed Mopra $^{13}$CO luminosity from the los velocity range -11 km s$^{-1}$ to 0 km s$^{-1}$ is $L_{^{13}CO}= 0.0022$ L$_\odot$. This corresponds then to roughly  $155$ M$_\odot$ of H$_2$ gas associated with the red CO component. In sum we find about 700 M$_\odot$ of evaporating or non-bound gas traced by CO. Adding the neutral gas traced by [C\,{\sc ii}], we find  $\sim 1700$ M$_\odot$  of neutral gas in the STO2 mapped region that is unbound to the Carina GMC, and possibly evaporating from it. }

This range can be compared to the mass in ionized HII gas that is evaporating from the GMC in the region mapped by STO2. In the previous subsection we derived that the ionized mass loss rate in the STO2 region is $\dot M_{evap} \simeq 5\times 10^{-3}$ M$_\odot$\ yr$^{-1}$. \citet{brooks01} and \citet{smith06} suggest an age for Tr 14 of about 1 Myr. Thus, roughly 5000 M$_\odot$ of ionized gas has evaporated from the surface of the GMC during this time and would mostly still lie inside the STO2 mapped area. { This appears to dominate the neutral evaporation by a factor of about 3. In summary, the mass loss rates we estimate lead to GMC lifetimes of 20 -- 30 Myr if they persist for the cloud lifetime, a result consistent with observational estimates of GMC lifetimes. }

\section{SUMMARY} \label{sec:con}

In this paper, we have presented a [C\,{\sc ii}] map of the Tr 14 region in the Carina Nebula Complex with an angular resolution of 48$''$ and a velocity resolution of 0.17 km s$^{-1}$. We investigated the distribution of the ISM phases and the kinematics of the Tr 14 region by comparing our [C\,{\sc ii}] spectral map to CO 1--0, HI 21cm, H$\alpha$, H$92\alpha$, [N\,{\sc ii}], [O\,{\sc i}],and dust continuum. The main results are as follows:

1. STO2, { a balloon-borne terahertz observatory}, has successfully observed [C\,{\sc ii}] in the Tr 14 region. Typical $rms$ of the [C\,{\sc ii}] map is 1.3 K in main-beam temperature at spatial and velocity resolutions of 48$''$ (0.5 pc at 2.3 kpc) and 0.17 km s$^{-1}$.

2. The [C\,{\sc ii}] emission is found in a broad region of the Tr 14 region (83\% and 53\% of the map area, 0.25$^\circ$ $\times$ 0.28$^\circ$, found to be above 5 K and 10 K, respectively.). The brightest [C\,{\sc ii}] emission is 370 K km s$^{-1}$ at Carina I-E. The [C\,{\sc ii}] emission shows good agreement with 8 $\mu$m PAH emission and warm dust emission at 160 $\mu$m rather than colder dust emission at 500 $\mu$m. Bright [C\,{\sc ii}] emission is mostly related with the CO clumps, while the [C\,{\sc ii}] emission near the center of the Tr 14 cluster (HII region) is relatively weak. This suggests that strong [C\,{\sc ii}] emission typically arises from PDRs and HII region close to the ionization fronts of nearby molecular clouds.

3. The channel maps of [C\,{\sc ii}] and CO 1--0 show that most of the CO emission is confined within a velocity range of -32 km s$^{-1}$ to -5 km s$^{-1}$. The bright [C\,{\sc ii}] emission is typically correlated with CO clouds/clumps but the intensity peaks of the [C\,{\sc ii}] emission are often displaced from the peak intensity of the CO 1--0 emission and located at the outskirts of the CO clouds/clumps in position-position-velocity. This indicates that the bright [C\,{\sc ii}] emission traces the PDRs and ionization fronts of the molecular clouds/clumps.

4. Absorption of the HI 21cm emission is in a good agreement with the CO clouds/clumps and the Keyhole Nebula, a CO-dark molecular cloud, in position-position space. In velocity space, we find that the absorption peaks of the HI 21cm emission are typically offset by 3 km s$^{-1}$ to 15 km s$^{-1}$ in LSR velocity from the intensity peaks of the CO 1--0 and [C\,{\sc ii}] emission, which is similar to the speed of the dispersing cloud by photoevaporation and radiation stripping in analytic and numerical studies. The rough estimate of evaporated mass is 5000 M$_\odot$ in the Tr 14 region.

5. We have modeled 10 PDRs from 10 spectral features along 7 {\it los}, using our PDR code to find good fits to the CO, [C\,{\sc i}], [O\,{\sc i}] and [C\,{\sc ii}] lines and the IR continuum. These fits provide the thermal pressure $P_{th}$ and $G_0$ for each PDR. In general, these pairs lie along a theoretically predicted relation of $P_{th} \propto G_0^{0.75}$ if thermal pressure dominates in the PDR. The empirical relation of P$_{th}$ to G$_0$ found by Meudon best fit PDRs in \citet{wu18}, $P_{th} \propto G_0^{0.9}$ lies close to the theoretical relation and our best fits, but factors of 2--4 above them. This could be due to different physical conditions in the regions mapped by \citet{wu18} (for example, small clumps or localized UV sources), or by differences in the PDR codes used to find best fits. {  In the 10 regions modeled, four regions had the PDR dominating production of the [C\,{\sc ii}], whereas six regions had the HII region dominating [C\,{\sc ii}] production. Averaging over all 10 regions, we found a mass proportion of 1:4.1:5.6 for the atomic:dark:molecular(CO) gas. The dark gas mass in these regions is comparable to the total mass traced by CO. Similarly, we summed the [C\,{\sc ii}] emission in the 10 regions modeled to get an average ratio in these regions of 3.7 for the [C\,{\sc ii}] arising from the HII region to the [C\,{\sc ii}] arising from the PDR. This ratio is not representative of the entire GMC since our 10 regions focus on los that penetrate the large blister HII region around Tr 14. }

6. Finally, combining the distribution of the [C\,{\sc ii}], CO 1–-0, H$\alpha$92, HI 21cm, and optical recombination line emission in PPV space, we find that the three dimensional morphology of the CNC is consistent with one side of numerous blister HII regions expanding freely toward us, with each acting in a manner analogous to a Champagne flow. The far-side of the HII regions seem to be confined by either CO or HI clouds, showing significant HI emission at LSR velocity larger than 0 km s$^{-1}$. CO clouds/clumps are found at the boundary of the HII region or embedded within the HII region. The HI absorption is correlated with the [C\,{\sc ii}] and CO emission in position-position space but is displaced in velocity space suggesting that the PDR regions highlighted by the strong [C\,{\sc ii}] emission have evaporating surface or radiation stripping with velocity around 3 -- 15 km s$^{-1}$. Lifetimes of the GMCs roughly estimated using the cloud dispersal rate measured in  [C\,{\sc ii}] and CO are 20 -- 30 Myr depending on the assumption of the optical depth of the lines.

\acknowledgments
We thank the anonymous referee for providing constructive comments that have improved the contents of this paper. We thank Ronin Wu for sharing PACS observations. STO2 is a multi-institutional effort funded by the National Aeronautics and Space Administration (NASA) through the ROSES-2012 program under grant NNX14AD58G. This research was carried out in part at the Jet Propulsion Laboratory, California Institute of Technology, under a contract with the NASA. YS acknowledges support from the NASA postdoctoral program. M.G.W. was supported in part by NSF Grant AST-1411827. CK was supported through NSF grant AST-1410896. Kristina Davis is currently supported by an NSF Astronomy and Astrophysics Postdoctoral Fellowship under award AST-1801983.

\appendix

\section{Intensity Calibration of STO2 Data Using ISO Observations}\label{app:cal}

{ We review a method for calibrating the intensity of the STO2 data. A conventional method of calibrating intensity for sub-millimeter/millimeter data is to use planets. The limited visibility of planets from the Antarctic during the STO2 flight coupled with limited broadband instrument stability made such a calibration approach problematic. \citet{oberst11} have mapped Tr 14 and Carina Nebulae in 158 $\mu$m [C\,{\sc ii}] using {\it ISO}. We use their data to verify the integrated intensity of our STO2 data toward Tr 14 \& Carina I.}

{ The [C\,{\sc ii}] data of \citet{oberst11} are in units of W m$^{-2}$ sr$^{-1}$, while the STO2 data are in K km s$^{-1}$. We convert the STO2 data to intensity in units of W m$^{-2}$ sr$^{-1}$. Using the Rayleigh-Jeans law, the integrated intensity of an STO2 spectrum is
\begin{eqnarray}
I_{\rm int, STO2} = \int I d\nu
  = \int {2k_B \over \lambda^2}T_A d\nu, \label{RJflux}
\end{eqnarray}
where $I_{\rm int}$ is the measured integrated intensity, $T_A$ is antenna temperature, $\nu$ is frequency, $\lambda$ is wavelength, and $k_B$ is the Boltzmann constant. We can write equation (\ref{RJflux}) in terms of velocity as
\begin{eqnarray}
I_{\rm int, STO2} = {2k_B \over \lambda^3} \int T_A dv, \label{RJflux2}
\end{eqnarray}
where $v$ is velocity. Considering a uniform source that fills only the main beam of the antenna, we get the true integrated intensity, $I^*_{\rm int}$, from $T_A = \eta_{\rm MB} T_{\rm MB}$, 
\begin{eqnarray}
I^*_{\rm int, STO2} = {I_{\rm int, STO2} \over \eta_{\rm MB}}, \label{int_true}
\end{eqnarray}
where $\eta_{\rm MB}$ is the main beam efficiency converting $T_{\rm MB}$ to $T_A$. For a symmetric Cassegrain antenna such as that of the STO2, a typical value of $\eta_{\rm MB}$ is 0.7.}

{ The integrated intensities in \citet{oberst11} are calculated using the equation
\begin{eqnarray}
I^*_{\rm int, ISO} = {\int F_\nu d\nu \over \Omega_{ISO}}, \label{obersteq2}
\end{eqnarray}
where $F_\nu$ is the flux of the ISO spectral line (W m$^{-2}$), and $\Omega_{ISO}$ is the solid angle of the ISO beam. We assume here that the ISO beam has a circular Gaussian form and that a spectral line has a Gaussian profile. Then, the solid angle is related to the FWHM of the Gaussian beam, $\theta_{\rm FWHM}$, by
\begin{eqnarray}
\Omega = {\pi\over4\ln2}\theta_{\rm FWHM}^2. \label{omega}
\end{eqnarray}
The integrated flux is related to the peak flux, $F_{\nu,peak}$, and the FWHM of the Gaussian fit to the ISO spectral line, $\Delta\nu$, by
\begin{eqnarray}
\int F_\nu d\nu = \sqrt{\pi \over 4\ln2} F_{\nu,peak} \Delta\nu.  \label{int_profile}
\end{eqnarray}
Using equations (\ref{omega}) and (\ref{int_profile}), we can rewrite equation (\ref{obersteq2}) as
\begin{eqnarray}
I^*_{\rm int, ISO} = \sqrt{4\ln2\over\pi}{F_{\nu,peak} \over\theta_{\rm FWHM}^2} \Delta\nu. \label{obersteq3}
\end{eqnarray}
Note that the above equation is not the same as equation (2) in \citet{oberst11}, which contains typographical errors: the above equation should replace equation (2) in \citet{oberst11}.}

\begin{deluxetable}{c c c c c c}[t]
\tablecolumns{6}
\tablewidth{0pt}
\tablecaption{Integrated intensities of [C\,{\sc ii}] at 12 commonly-observed points using the STO2 and the ISO.\label{table:cal}}
\tablehead{
\colhead{Points} & \colhead{$l$} & \colhead{$b$} & \colhead{$I^*_{\rm STO2}$} & \colhead{$I^*_{\rm ISO}$} & \colhead{$R_c$} \\
\colhead{ } & \colhead{Degree} & \colhead{Degree} & \colhead{10$^{-4}$ erg/s/cm$^2$/sr} & \colhead{10$^{-4}$ erg/s/cm$^2$/sr} & \colhead{ } 
}
\startdata
1  & 287.256 & -0.535 &  4.0 &  4.1 & 0.98 \\
2  & 287.256 & -0.585 &  8.1 &  9.2 & 0.88 \\
3  & 287.405 & -0.586 & 11.2 &  7.4 & 1.52 \\
4  & 287.405 & -0.536 &  8.8 &  9.2 & 0.96 \\
5  & 287.255 & -0.687 &  8.3 &  8.2 & 1.02 \\
6  & 287.254 & -0.634 &  8.0 &  7.7 & 1.04 \\
7  & 287.405 & -0.686 &  9.1 &  7.3 & 1.2  \\
8  & 287.405 & -0.637 & 15.5 &  7.3 & 2.1  \\
9  & 287.255 & -0.687 &  8.8 &  8.6 & 0.96 \\
10 & 287.305 & -0.686 &  6.6 & 11.5 & 0.59 \\
11 & 287.355 & -0.686 & 11.5 & 16.3 & 0.71 \\
12 & 287.405 & -0.686 &  9.1 &  9.6 & 0.94 \\
\enddata
\end{deluxetable}

{ We were able to compare 12 points from the Table 3 in \citet{oberst11} to our STO2 map. For a proper comparison, we convolved the STO2 map to have the same beam size as the ISO beam size (70.1$''$). We excluded the points that are at the edge of the STO2 map. The Galactic coordinates of 12 points, the intensities, and the intensity ratios are given in the Table 1. The intensity ratio of STO2 compared to ISO is
\begin{eqnarray}
R_c = {I^*_{\rm int, STO2}\over I^*_{\rm int, ISO}}. \label{couple}
\end{eqnarray}
The intensity ratio should be unity if ISO and STO2 are correctly calibrated. We find that the intensity ratio in Table 1 varies somewhat. The median value of the intensity ratio is 0.97, and the standard deviation is 0.38. Both ISO and STO2 have uncertainties in the data reduction and calibration. \citet{oberst11} wrote that their uncertainty in the calibration is about 20\%. STO2 also has calibration uncertainties. Particularly, some of the STO2 data have quite large fringes (baseline ripple). We found that there is $>$30\% uncertainty in the peak intensity depending on the reduction method. Propagating the uncertainties from both ISO and STO2, the uncertainty of the ratio is 36\%, which is similar to the standard deviation of the ratio in Table \ref{table:cal}. }

{ From having the 1$-\sigma$ uncertainty of $R_c$ of $^{0.2}_{-0.1}$, we conclude that the STO2 data corrected by the main beam efficiency are consistent with the ISO intensities. A beam efficiency of 0.7 should be employed for the STO2 data, and no further adjustment is included at this time.  }

\section{Channel Maps}\label{app:channel}

\begin{figure*}[h!]
\centering
\includegraphics[angle=0,scale=0.315]{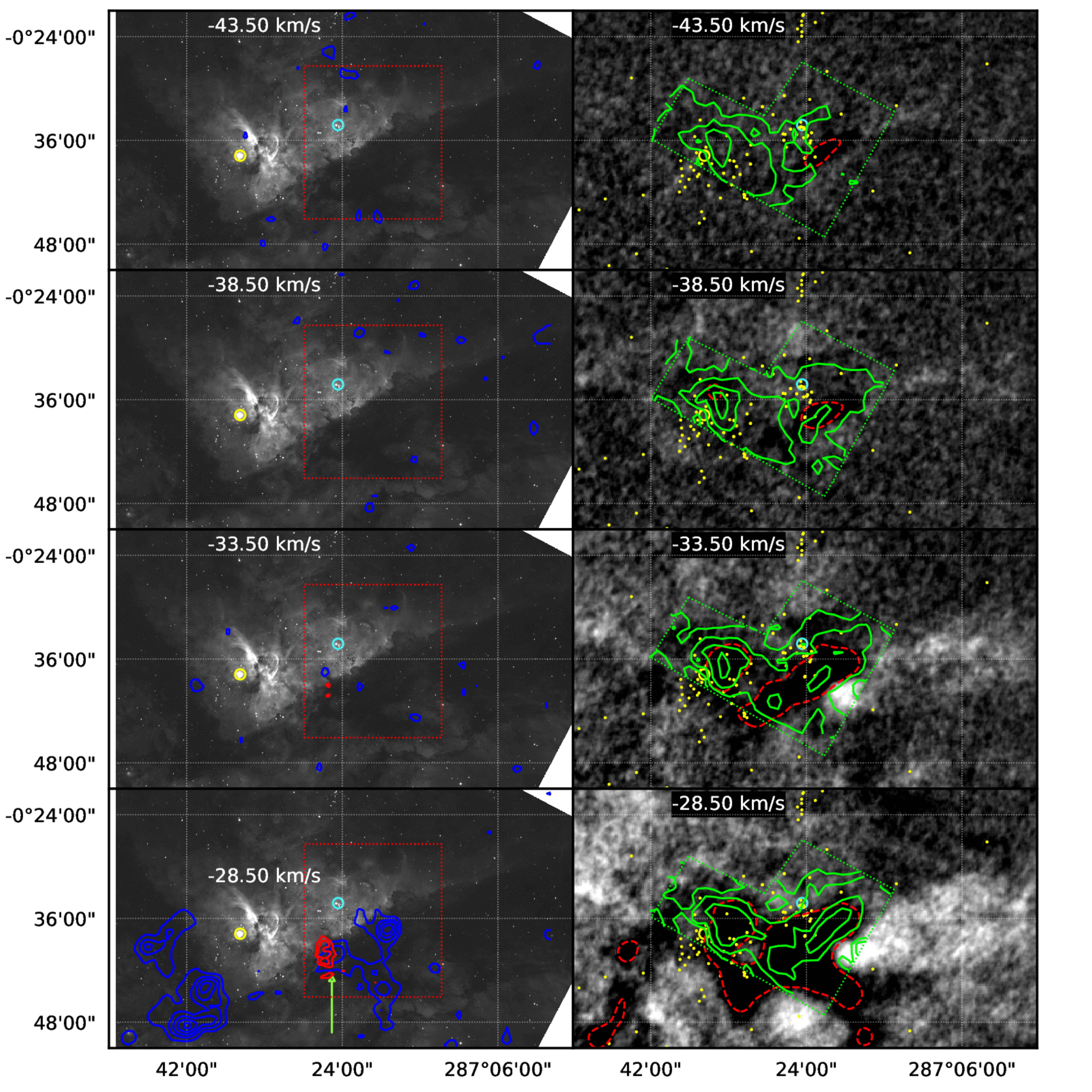}
\caption{Channel maps of [C\,{\sc ii}] 158 $\mu$m obtained with STO2 (red contours), $^{12}$CO observed with Mopra (blue contours, \citealt{rebolledo16}), H92$\alpha$ (green contours, \citep{horiuchi12}), and HI 21cm (grayscale background in the right column, \citealt{rebolledo17}). The grayscale background in the left column is an H$\alpha$ image \citep{smith06}. The [C\,{\sc ii}] contours start at 7.5 K and increase in 2.5 K increments. The $^{12}$CO contours start at 1.5 K and increase in 1.5 K increments. The H92$\alpha$ contours are at 0.25, 0.5, 1, 1.5, 2, and 2.5 K. The red and green boxes with the dotted lines are the areas mapped in [C\,{\sc ii}] and H92$\alpha$, respectively. The red dashed contours are at an antenna temperature of -30 K in HI 21 cm to show the absorption cavity. The large yellow and cyan circles indicate $\eta$ Carinae and the center of the Trumpler 14 cluster. The yellow dots denote O- and B-type stars \citep{alexander17}. The green arrow denotes the [C\,{\sc ii}] emission associated with a dense CO clump.}
\label{fig_chan1}
\end{figure*}

\begin{figure*}[h!]
\centering
\includegraphics[angle=0,scale=0.315]{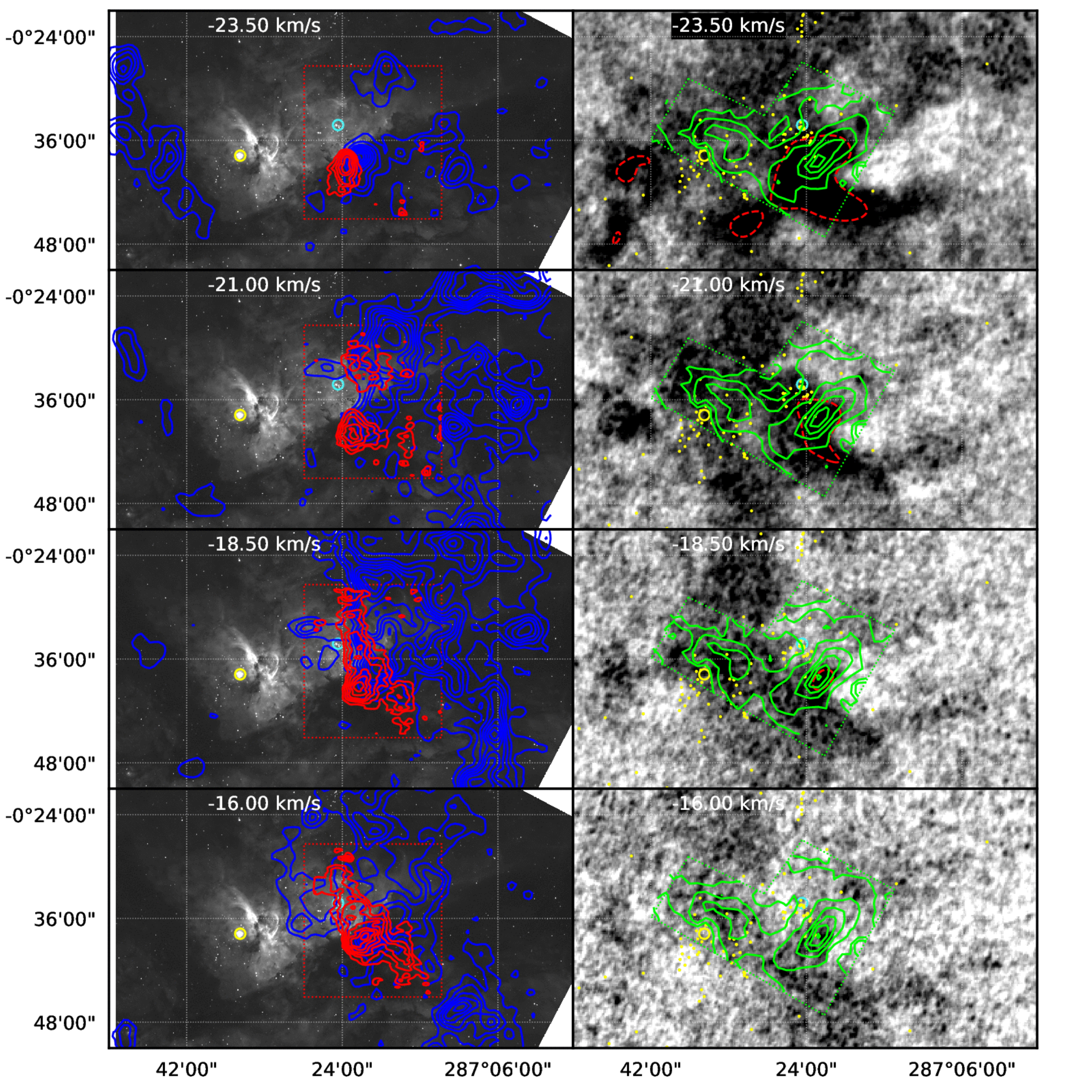}
\caption{Channel maps continued from Figure \ref{fig_chan1}.}
\label{fig_chan2}
\end{figure*}

\begin{figure*}[h!]
\centering
\includegraphics[angle=0,scale=0.315]{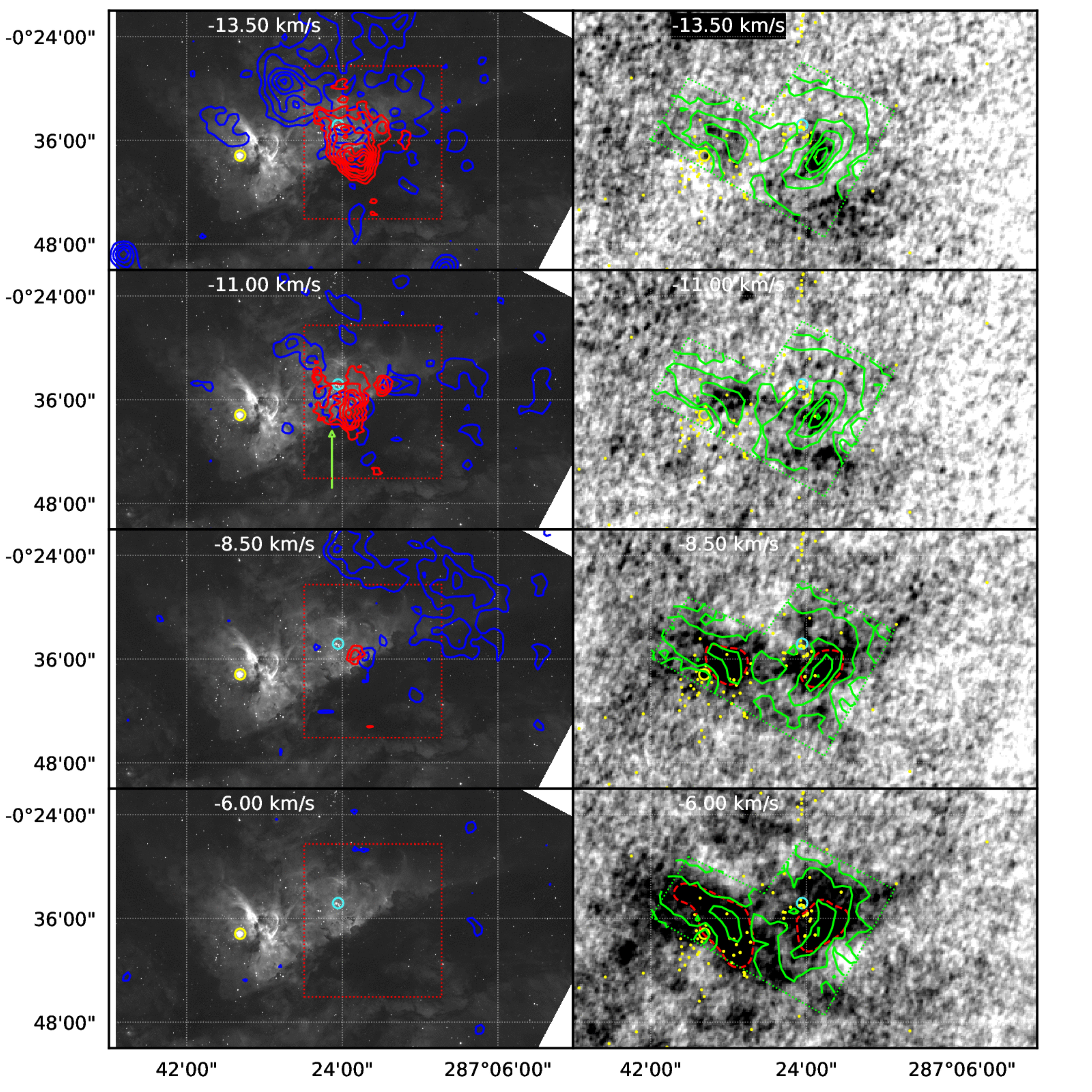}
\caption{Channel maps continued from Figure \ref{fig_chan2}.}
\label{fig_chan3}
\end{figure*}

\begin{figure*}[h!]
\centering
\includegraphics[angle=0,scale=0.315]{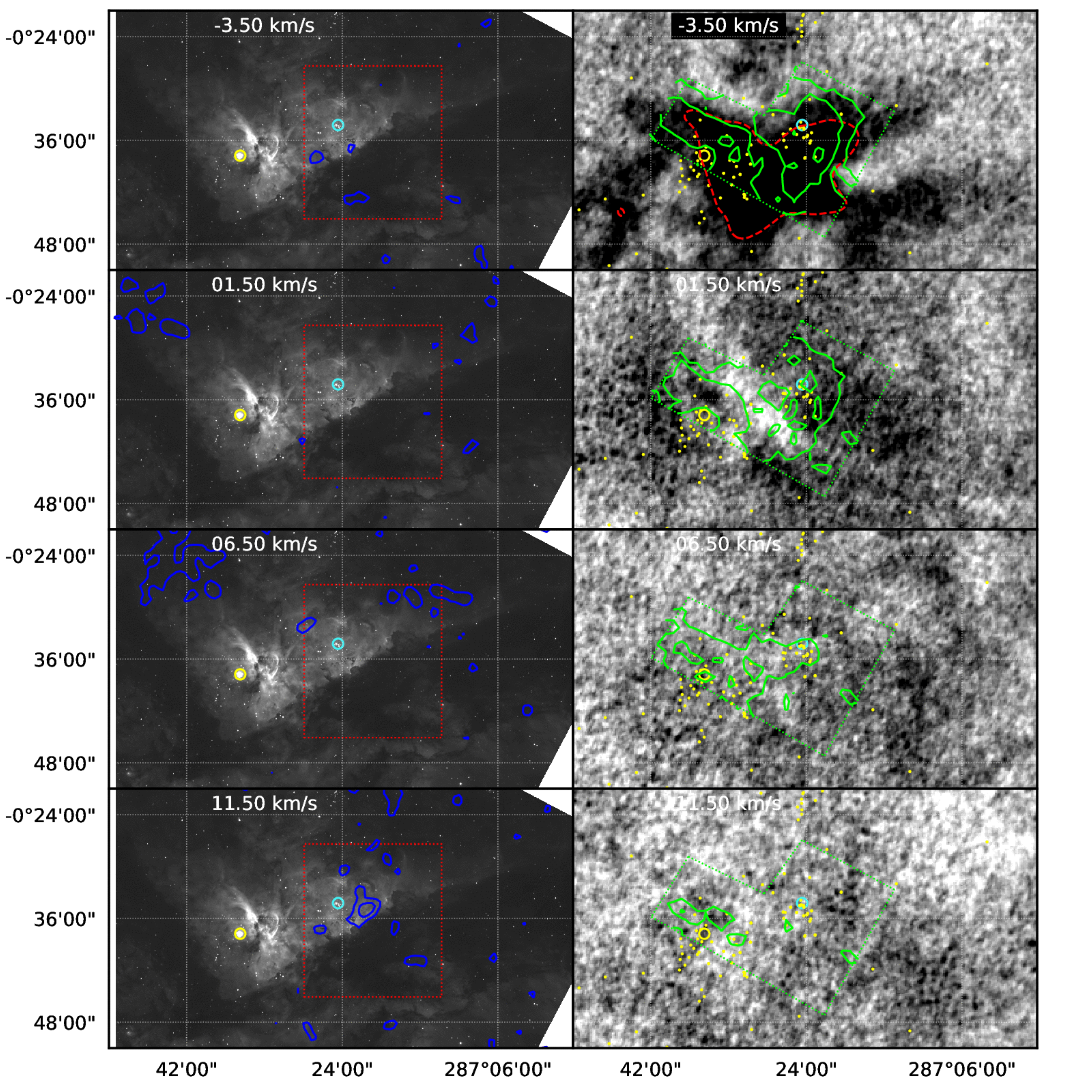}
\caption{Channel maps continued from Figure \ref{fig_chan3}.}
\label{fig_chan4}
\end{figure*}

{ We show channel maps of [C\,{\sc ii}], $^{12}$CO 1--0, H92$\alpha$, and HI 21 cm from -43.5 km s$^{-1}$ to 11.5 km s$^{-1}$ (Figures \ref{fig_chan1} -- \ref{fig_chan4}). These channel maps show details of the relative distribution of the different ISM phases in PPV space.} 


\clearpage

\bibliographystyle{aasjournal}
\bibliography{Tr14_ref}

\end{document}